\PassOptionsToPackage{hyphens}{url}
\documentclass[a4paper, reprint, preprintnumbers, amsmath, amssymb, superscriptaddress, nofootinbib, aps, pra, floatfix]{revtex4-2}

\usepackage[english]{babel}
\usepackage[T1]{fontenc}
\usepackage[utf8]{inputenc}

\usepackage{amsmath}
\usepackage{amsfonts}
\usepackage{amssymb}
\usepackage[
]{graphicx}
\usepackage[colorinlistoftodos]{todonotes}
\usepackage[utf8]{inputenc}
\usepackage{bbold}
\usepackage{cancel}
\usepackage{comment}
\usepackage{mathtools}
\usepackage{hyperref}
\usepackage{physics}
\usepackage{hyperref}

\begin{document}

\title{Asymmetric node placement in fiber-based quantum networks}
\date{\today}

\author{Guus Avis}
\email{guusavis@hotmail.com}
\affiliation{QuTech, Delft University of Technology and TNO, Lorentzweg 1, 2628 CJ Delft, The Netherlands}
\affiliation{Quantum Computer Science, EEMCS, Delft University of Technology, Lorentzweg 1, 2628 CJ Delft, The Netherlands}
\affiliation{Kavli Institute of Nanoscience, Delft University of Technology, Lorentzweg 1, 2628 CJ Delft, The Netherlands}

\author{Robert Knegjens}
\affiliation{QuTech, Delft University of Technology and TNO, Lorentzweg 1, 2628 CJ Delft, The Netherlands}

\author{Anders S. Sørensen}
\affiliation{Center for Hybrid Quantum Networks (Hy-Q), Niels Bohr Institute, University of Copenhagen, Blegdamsvej 17, DK-2100 Copenhagen Ø, Denmark}

\author{Stephanie Wehner}
\email{s.d.c.wehner@tudelft.nl}
\affiliation{QuTech, Delft University of Technology and TNO, Lorentzweg 1, 2628 CJ Delft, The Netherlands}
\affiliation{Quantum Computer Science, EEMCS, Delft University of Technology, Lorentzweg 1, 2628 CJ Delft, The Netherlands}
\affiliation{Kavli Institute of Nanoscience, Delft University of Technology, Lorentzweg 1, 2628 CJ Delft, The Netherlands}

\begin{abstract}

Restrictions imposed by existing infrastructure can make it hard to ensure an even spacing between the nodes of future fiber-based quantum networks.
We here investigate the negative effects of asymmetric node placement by considering separately the placement of midpoint stations required for heralded entanglement generation, as well as of processing-node quantum repeaters in a chain.
For midpoint stations, we describe the effect asymmetry has on the time required to perform one entangling attempt, the success probability of such attempts, and the fidelity of the entangled states created.
This includes accounting for the effects of chromatic dispersion on photon indistinguishability.
For quantum-repeater chains we numerically investigate how uneven spacing between repeater nodes leads to bottlenecks, thereby increasing both the waiting time and the time states are stored in noisy quantum memory.
We find that while the time required to perform one entangling attempt may increase linearly
with the midpoint's asymmetry,
the success probability and fidelity of heralded entanglement generation
and the distribution time and error rate for repeater chains
all have vanishing first derivatives with respect to the amount of asymmetry.
This suggests resilience of quantum-network performance against small amounts of asymmetry.

\end{abstract}

\maketitle

\section{Introduction} \label{sec:introduction}

The quantum internet promises the distribution of quantum entanglement between any two points on the planet \cite{wehner2018}.
Entanglement can be a valuable resource that enables a variety of applications in domains such as quantum cryptography \cite{bennett2014, bennett2014, ekert1991, pirandola2020, leichtle2021}, distributed quantum computing \cite{grover1997, cirac1999a, caleffi2022} and quantum sensing \cite{komar2014, guo2020, gottesman2012}.
A major outstanding challenge towards the construction of large-scale ground-based quantum networks is the fact that entangling rates over optical fiber decline exponentially with the length of the fiber due to attenuation losses.
While classical amplification strategies are not viable, a special class of devices called quantum repeaters can be used to reach high entangling rates over long optical fibers \cite{briegel1998, dur1999}.
By using quantum repeaters as intermediate nodes a fiber is divided into segments over which entanglement can be distributed more efficiently than over the full fiber length.
Although various proposals for quantum repeaters exist \cite{azuma2022, inside_quantum_repeaters}, only proof-of-concept experiments have been realized so far \cite{bhaskar2020, langenfeld2021a}.
\\

In order to build a quantum network, decisions need to be made on where its nodes are positioned and how they are connected.
These nodes include the end nodes of the network, quantum repeaters and potentially midpoint stations as required by some entanglement-generation protocols (as depicted in Figure \ref{fig:midpoint_asymmetry}) \cite{cabrillo1999, barrett2005}.
For the midpoint stations, and for quantum repeaters running at least some specific types of protocols (such as the one investigated in this paper),
optimal network performance requires the nodes to be positioned \textit{symmetrically}~\cite{rozpedek2018, rozpedek2019, pathumsoot2021}.
That is, with an internode spacing that is the same between all neighboring nodes.
To understand why symmetric placement of repeater nodes can be favourable to the performance, consider the following.
For quantum repeaters in a chain, the end-to-end capacity for generating entanglement is equal to the minimal capacity over all pairs of neighboring nodes in the chain \cite{pirandola2019}.
This minimal capacity is maximized by a symmetric placement of repeater nodes, and hence a symmetric placement optimizes the end-to-end capacity.
We note however that there exist specific repeater protocols that do perform best, according to specific performance metrics, under asymmetric node placement \cite{jiang2007, luong2016}.
The suboptimal capacity of such a node placement suggests that improvements on the protocols could perhaps result in a symmetric placement being optimal again.
This is demonstrated by Ref. \cite{rozpedek2018}, where it is shown that the advantage found in Ref. \cite{luong2016} vanishes when one further optimizes the protocol.
\\

Symmetric placement of nodes in a quantum network may not always be possible.
For instance, if a quantum network is built using existing infrastructure this restricts the freedom in choosing the locations of the nodes \cite{rabbie2022}.
Therefore, in this paper, we address the question of how asymmetric node placement affects the performance of a quantum network.
We do so by investigating two separate aspects of asymmetric quantum networks:
first, we consider asymmetric placement of midpoint stations and examine how entanglement generation between two neighboring nodes is affected (Section \ref{sec:midpoint}).
We identify three independent effects, namely on the cycle time of entanglement generation (Section \ref{sec:cycle_time}, see the beginning of Section \ref{sec:midpoint} for a definition), on the success probability of entanglement generation and the fidelity of generated entanglement through the introduction of imbalanced losses (Section \ref{sec:imbalanced_losses}), and on the photon indistinguishability through chromatic dispersion (Section \ref{sec:dispersion}).
Second, we consider asymmetric placement of quantum repeaters in a chain (Section \ref{sec:repeater}).
Here, we focus specifically on processing-node repeaters executing a SWAP-ASAP protocol (as explained in Section \ref{sec:repeater} and studied in, e.g., Refs. \cite{coopmans2021, inesta2023}).
Notably, the results presented in this paper indicate robustness against small amounts of asymmetry.
For asymmetry in the placement of the midpoint station, we find that both the success probability and fidelity have a vanishing first derivative with respect to how asymmetrically the midpoint is positioned (granted that the photons are shaped such that the effects of chromatic dispersion are negligible).
However, the cycle time increases linearly with the asymmetry in case the time required to exchange signals between neighboring nodes is the limiting factor (it may be independent of asymmetry if this is not the case).
Similarly, we find that both the entangling rate and error rate in a SWAP-ASAP repeater chain have vanishing first derivatives with respect to how asymmetrically the repeater nodes are positioned.
This robustness suggests that, when designing a quantum network, nodes do not need to be placed exactly symmetrically.
It furthermore suggests that the effects of constraints on node locations imposed by existing infrastructure on network performance may not be too severe.

\section{Asymmetry in midpoint placement}
\label{sec:midpoint}

Two popular methods for the creation of entanglement between neighboring nodes in a quantum network are single-click heralded entanglement generation \cite{cabrillo1999, humphreys2018, pompili2021} and double-click heralded entanglement generation (also known as the Barrett-Kok protocol) \cite{barrett2005, bernien2013, hensen2015, stephenson2020, krutyanskiy2023}.
In both of these protocols, time is slotted.
In each time slot, the nodes perform a single attempt at entanglement generation.
Such an attempt consists of both nodes sending a photon entangled with a local qubit to a midpoint station, where the photons are interfered and measured.
The midpoint then sends a message to the end nodes containing the measurement outcome.
Depending on the measurement outcome, the attempt is declared either a success or a failure.
The probability that it is declared a success is called the success probability and denoted by $P_\text{succ}$.
The duration of each time slot (i.e., the time required to perform one attempt) is called the cycle time and denoted by $T_\text{cycle}$.
The (average) rate at which successes occur is then given by
\begin{equation}
\label{eq:rate}
R = \frac{P_\text{succ}}{T_\text{cycle}}.
\end{equation}
After a successful attempt a state $\rho$ is shared by the two neighboring nodes.
Ideally, the state $\rho$ is some pure maximally-entangled target state $\ketbra \phi$.
However, due to noise, $\rho$ will instead be a mixed state with fidelity
\begin{equation}
F = \expval{\rho}{\phi}.
\end{equation}
We will use the success probability $P_\text{succ}$, the cycle time $T_\text{cycle}$ and the fidelity $F$ as performance metrics for heralded entanglement generation.
\\

\begin{figure}
    \includegraphics[width=\linewidth]{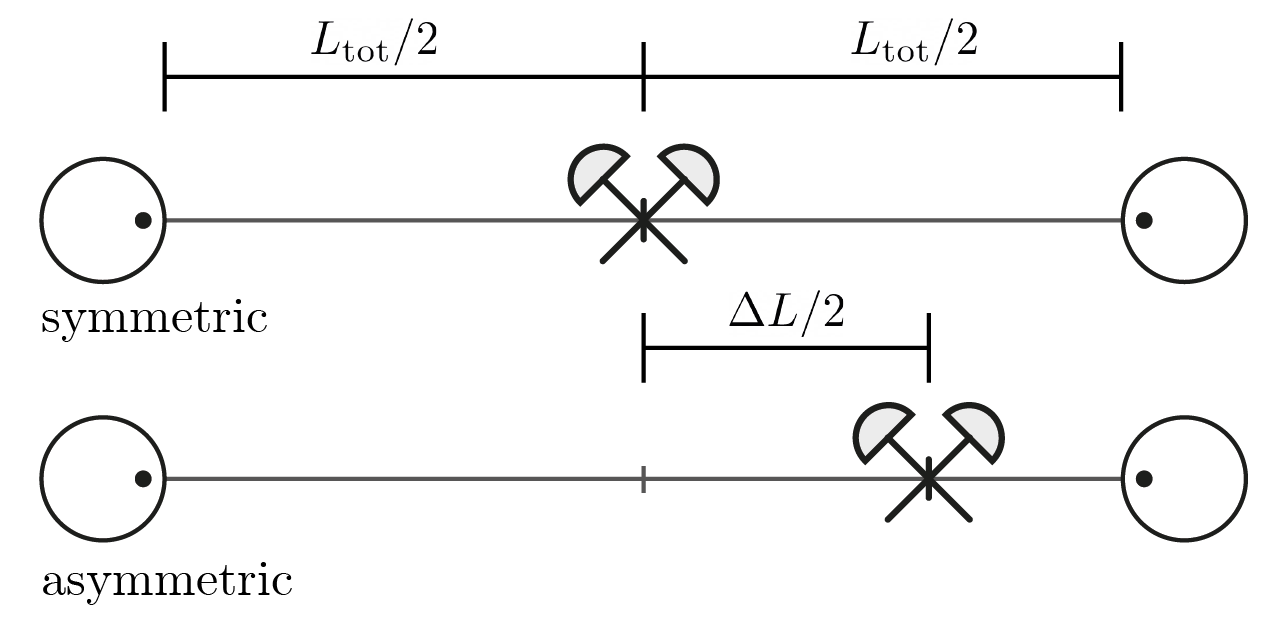}
    \caption[
        Midpoint asymmetry.
    ]
    {
        Symmetric and asymmetric positioning of a midpoint station for heralded entanglement generation.
        The magnitude of the parameter $\Delta L$ is a measure for how large the asymmetry is.
        $\Delta L$ and $L_\text{tot}$ are defined in Equation \eqref{eq:L}.
    }
    \label{fig:midpoint_asymmetry}
\end{figure}

In this section we study the effect of displacing the midpoint station from the exact center between the nodes (as illustrated in Figure \ref{fig:midpoint_asymmetry}) on our performance metrics.
We do so by separately examining the effect on the cycle time, the effect that the resulting imbalanced losses have on the success probability and the fidelity, and the effect on the photon indistinguishability (which in turn affects primarily the fidelity but also the success probability).
In order to do so we first need a method for quantifying how far the midpoint has been displaced.
To that end, let the fiber distance between the midpoint station and the left-hand (right-hand) node be denoted $L_\text{left}$ ($L_\text{right}$).
Then we define
\begin{equation}
\label{eq:L}
\begin{aligned}
\Delta L &= L_\text{left} - L_\text{right}, \\
L_\text{tot} &= L_\text{left} + L_\text{right}.
\end{aligned}
\end{equation}
The parameter $\Delta L$ is then a measure of the amount of asymmetry, as shown in Figure \ref{fig:midpoint_asymmetry}.
As we will show below, the effects of asymmetric midpoint placement on the cycle time, success probability and fidelity are all quantified by $|\Delta L|$.

\subsection{Cycle time}
\label{sec:cycle_time}

First we consider the effect of asymmetric midpoint placement on the cycle time of the entanglement-generation protocol between neighboring nodes.
During each cycle both nodes need to emit entangled photons that reach the midpoint station simultaneously.
Then the midpoint station sends a message with the measurement result back to each of the nodes.
Assuming both the entangled photons and the messages travel with the same velocity $c$, the cycle time at least includes the communication time between the midpoint station and the node that is furthest away.
That is, $T_\text{cycle} \geq \tfrac 2 c \max(L_\text{left}, L_\text{right})$.
This can be rewritten as
\begin{equation}
\label{eq:cycle_time}
T_\text{cycle} \geq \frac 1 c (L_\text{tot} + |\Delta L|).
\end{equation}
When the cycle time is limited only by the speed-of-light communication delay the cycle time will be exactly equal to the right-hand side of the equation.
However, we note that in practice the cycle time is often much longer (see, e.g., Refs \cite{pompili2021, krutyanskiy2023}), for example due to local operations or the limited rate at which entangled photons can be emitted.
In that regime, the cycle time may be independent of $\Delta L$ until the asymmetry becomes so large that the communication delays are again the limiting factor.

\subsection{Imbalanced losses}
\label{sec:imbalanced_losses}

As attenuation loss in fiber scales exponentially with the length of the fiber,
having a midpoint station that is off center will result in an imbalance between the losses encountered by the photons.
To be more precise, let $P_0$ be the probability that when a node attempts photon emission, this photon is emitted successfully, couples successfully to fiber, and is then successfully detected at a detector at the end of the fiber, given that the fiber has length zero.
Then, the probability that photon emission at the left node leads to photon detection at the midpoint station is given by 
\begin{equation}
P_\text{left} = P_0 e^{- \frac {L_\text{left}} {L_\text{att}}},
\end{equation}
where $L_\text{att}$ is the attenuation coefficient of the fiber.
The same equation holds for $P_\text{right}$, but with $L_\text{left}$ replaced by $L_\text{right}$.
In an asymmetric setup we will have $P_\text{left} \neq P_\text{right}$, which is what we mean by imbalanced losses.
This can affect both the success probability $P_\text{succ}$ and the fidelity $F$ of heralded entanglement generation.
\\

For both the single- and double-click protocol, expressions for $P_\text{succ}$ and $F$ in terms of, among other parameters, $P_\text{left}$ and $P_\text{right}$ can be found in Ref. \cite{avis2022a}.
In order to make the effect of imbalanced losses explicit in these expressions we introduce the parameters
\begin{equation}
\begin{aligned}
\label{eq:p_tot_p_sum}
P_\text{tot} &\equiv P_\text{left} P_\text{right} = P_0^2 e^{- \frac {L_\text{tot}} {L_\text{att}}},\\
P_\text{sum} &\equiv P_\text{left} + P_\text{right} = 2 \sqrt{P_\text{tot}} \cosh\left(\frac{|\Delta L|}{2L_\text{att}}\right).
\end{aligned}
\end{equation}
Nontrivially, we find that for both protocols (to leading order, as discussed below) we can eliminate $P_\text{left}$ and $P_\text{right}$ completely from the expressions for $P_\text{succ}$ and $F$ in favour of $P_\text{tot}$ and $P_\text{sum}$.
The effect of imbalanced losses is then captured entirely by the dependence of $P_\text{sum}$ on $\Delta L$.
We discuss the resulting expressions and their implications for the single- and double-click protocol separately below.
\\

In the double-click protocol, both nodes emit a photon.
The mode that the photon is emitted in (e.g., its polarization) is entangled with the state of the emitter, and entanglement between the emitters is heralded when both photons are detected in different modes at the midpoint station after interfering on a beam splitter.
By eliminating $P_\text{left}$ and $P_\text{right}$ as described above we find (see Appendix \ref{app:single_double_click})
\begin{equation}
\label{eq:double_click}
\begin{aligned}
P_\text{succ, 2click} =& d_1 + 2  p_\text{dc} P_\text{sum} + \mathcal O (p_\text{dc}^2), \\
F_\text{2click} =& d_2 - d_3 p_\text{dc} P_\text{sum} + \mathcal O(p_\text{dc}^2).
\end{aligned}
\end{equation}
The parameters $d_1$, $d_2$, and $d_3$ have no direct dependence on $\Delta L$ and are given by
\begin{equation}
\begin{aligned}
d_1 =& \frac 1 2 P_\text{tot} -  p_\text{dc} \left(4 + r  - \frac 1 2 (2 - r) (1 + V)\right)P_\text{tot},\\
d_2 =& \left(\frac 1 2 q_\text{em} (1 + V) + \frac 1 4 (1 - q_\text{em})\right) \left(1 + 8 p_\text{dc} \right)\\
&- \frac 1 2 (2 - r) q_\text{em} p_\text{dc} (1 + V)^2,\\
d_3 =& \frac{q_\text{em}}{P_\text{tot}} (2 V + 1).
\end{aligned}
\end{equation}
Here, $p_\text{dc}$ denotes the detector dark-count probability.
The notation $\mathcal O(x^n)$ represents any terms that are of order $n$ in the parameter $x$.
As $p_\text{dc}$ is typically small, we have only included leading-order terms in Equations \eqref{eq:double_click} (the full expressions can be found in Appendix \ref{app:single_double_click}).
$V$ denotes the indistinguishability of the photons, which can itself depend on the asymmetry through the effect of chromatic dispersion as discussed in Section \ref{sec:dispersion}.
It is is assumed that the state shared between a node's emitter and the photon it emits is given by $\tfrac 1 3 (4F-1) \ketbra{\psi} + \tfrac 1 3 (1 - F) \mathbb 1$,
which has fidelity $F$ to the state $\ket \psi = \tfrac 1 {\sqrt 2} (\ket{00} + \ket{11})$ and where $F=F_\text{em, left}$ ($F=F_\text{em, right}$) for the left (right) node.
We then have
\begin{equation}
q_\text{em} = \frac 1 9 (4 F_\text{em, left} - 1) (4 F_\text{em, right} - 1).
\end{equation}
Finally,
\begin{equation}
r =
\begin{cases}
1 \quad \text{ for non-photon-number-resolving detectors,}\\
2 \quad \text{ for photon-number-resolving detectors.}
\end{cases}
\end{equation}
We see that when the dark-count probability is zero, the double-click protocol is not affected by imbalanced losses at all.
This is explained by the fact that the probability of both photons surviving their respective fiber segments is equal to the probability of a single photon surviving the full fiber length $L_\text{tot}$, which is not affected by asymmetry.
The reason why the protocol is affected in the presence of dark counts is that as the photon arrival probability on the longer leg becomes of the same order as the dark-count probability, the probability of falsely heralding successful entanglement becomes large.
This results both in an increased rate and a reduced fidelity.
\\

In the single-click protocol, both nodes also perform photon emission and send those photons to the midpoint station.
However, before emission starts, the emitter is prepared in an unbalanced superposition of a bright state from which photons can be emitted and a dark state from which emission is impossible.
How large the amplitude of the bright state is, is parameterized by the bright-state parameter $\alpha$.
As a result, after emission, the state shared by the emitter and the photon takes the form
\begin{equation}
\sqrt{1 - \alpha} \ket{\text{dark}} \ket{0} + \sqrt \alpha \ket{\text{bright}} \ket{1},
\end{equation}
where $\ket 0$ ($\ket 1$) indicates the absence (presence) of the photon.
An attempt is then considered a success in case only one photon is detected at the midpoint station (as opposed to two for the double-click protocol),
creating an entangled state that is a superposition of the left-node emitter being in the bright state but the right-node emitter in the dark state and vice versa.
However, in case both emitters are in the bright state but one of the emitted photons is lost, a success is heralded without the creation of an entangled state.
Therefore, even when the only imperfection in the system is fiber attenuation, the created entangled state is never pure.
The fidelity of the created entangled state will depend on the choice of $\alpha$;
when $\alpha$ is small, the relative probability that both nodes are found in the bright state is suppressed resulting in a good fidelity.
However, using a small $\alpha$ also results in a small success probability.
In case the midpoint is placed symmetrically and there are no imperfections but losses, for $\alpha \ll 1$, the success probability and fidelity can be approximated as $P_\text{succ} \approx 2 \alpha \sqrt{P_\text{tot}}$ and $F \approx 1 - \alpha$.
See, e.g., Ref. \cite{childress2005} for a further discussion of this effect.
Thus, choosing the value of $\alpha$ is a matter of trading off success probability and fidelity.
In an asymmetric setup it has been found that in case one wants to optimize the fidelity, the equation $\alpha_\text{left} P_\text{left} \approx \alpha_\text{right} P_\text{right} $ should be satisfied \cite{pompili2021}.
Therefore, we here assume the bright-state parameters are always chosen such that
\begin{equation}
\label{eq:single_click_q}
\alpha_\text{left} P_\text{left} = \alpha_\text{right} P_\text{right} \equiv q,
\end{equation}
where $q$ parameterizes the remaining degree of freedom.
As the bright-state parameter needs to be small in order to get a good fidelity, we will here present a result that is not only leading order in the dark-count probability but also in the parameter $q$.
Eliminating $\alpha_\text{left}$ and $\alpha_\text{right}$ in favour of $q$ and $P_\text{left}$ and $P_\text{right}$ in favour of $P_\text{tot}$ and $P_\text{sum}$ we find (see Appendix \ref{app:single_double_click})
\begin{equation}
\begin{aligned}
\label{eq:single_click}
P_\text{succ, 1click} =& 2q + 2 p_\text{dc} + \mathcal O(q^2, p_\text{dc}^2, qp_\text{dc}) , \\
F_\text{1click} =& s_1 - s_2 P_\text{sum} + \mathcal O(q^2, p_\text{dc}^2, qp_\text{dc}).
\end{aligned}
\end{equation}
Here, the parameters $s_1$ and $s_2$ are defined by
\begin{equation}
\begin{aligned}
s_1 =& \frac 1 2 (1 + \sqrt V) \frac q {q + p_\text{dc}} \Bigg(1 + q - (1 + r) p_\text{dc}\\
&+ \frac q {q+p_\text{dc}} \left[ r p_\text{dc} - \frac 1 4 (2 - r) (1 + V) q \right] \Bigg), \\
s_2 =& \frac 1 2 (1 + \sqrt V) \frac q {q + p_\text{dc}} \frac {1}{P_\text{tot}} \left( \frac 1 2 q - p_\text{dc} \right).
\end{aligned}
\end{equation}
Note that the success probability of the single-click scheme is not affected when $\Delta L$ is increased, as long as the bright-state parameters are chosen to keep $q$ constant.
This behaviour is not a consequence of the leading-order expansion.
It is shown in Appendix \ref{app:single_double_click} that the exact expression for the success probability has no direct dependence on the asymmetry either.
\\

\begin{figure}
    \includegraphics[width=\linewidth]{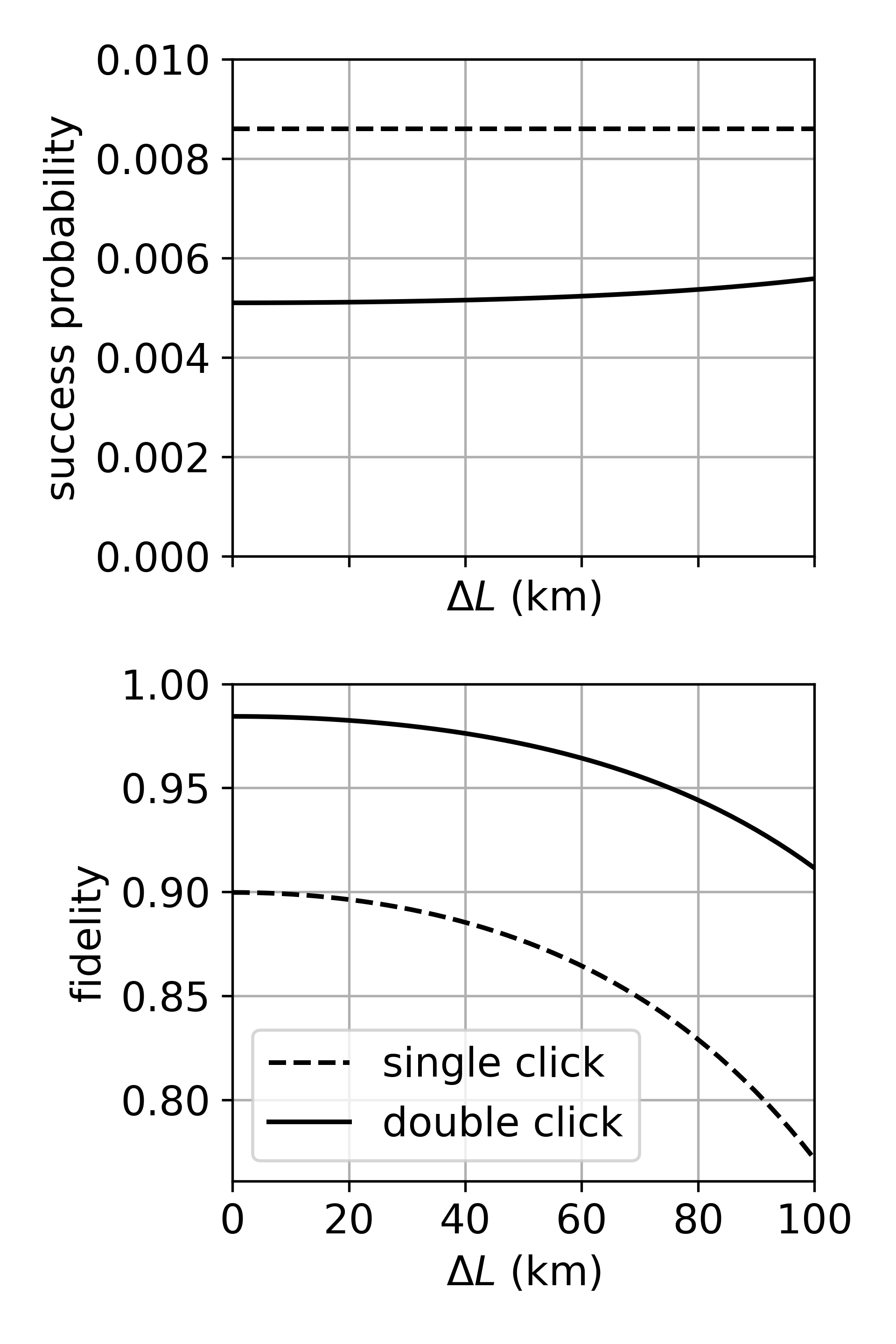}
    \caption[
        Effect of imbalanced losses on success probability and fidelity.
    ]
    {
        Leading-order results (presented in Equations \eqref{eq:double_click} and \eqref{eq:single_click}) for the probability that an entanglement-generation attempt is heralded as a success and the fidelity of entangled states created upon a heralded success of both the single-click and double-click protocol as a function of the difference in length between the two fibers connecting the midpoint station ($\Delta L$, defined in Equation \eqref{eq:L}).
        This figure has been created using the parameters $L_\text{tot} = 100$ km, $p_\text{dc} = 3 \cdot 10^{-4}$, $q=4 \cdot 10^{-3}$ and $L_\text{att} \approx 22$ km.
        Apart from attenuation losses and dark counts no imperfections have been included.
    }
    \label{fig:imbalance}
\end{figure}

The success probability and fidelity as a function of the asymmetry are shown in Figure \ref{fig:imbalance} for both protocols.
We see that in both cases imbalanced losses do not reduce the success probability.
Additionally the fidelity falls in a similar way for both cases, with a vanishing first derivative at $\Delta L = 0$.
The reason for this is that the hyperbolic cosine to which $P_\text{sum}$ is proportional (see Equations \eqref{eq:p_tot_p_sum}) has a vanishing first derivative at zero.
As a result, the success probability and fidelity are resilient against small amounts of asymmetry.
For instance, for the parameters considered in Figure \ref{fig:imbalance}, we see that the fidelity is still above 99\% of the value it attains for symmetric midpoint placement at $\Delta L = 30$ km.

\subsection{Photon indistinguishability}
\label{sec:dispersion}

Light waves traveling through optical fiber are subject to chromatic dispersion, meaning that different frequency components travel at different velocities.
As a result, when performing heralded entanglement generation, the photons that arrive at the midpoint station are shaped differently than the photons that are emitted by the nodes.
A key requirement for the creation of an entangled state through the interference and measurement of the photons is that the photons are indistinguishable, i.e., their wave packets need to be identical and arrive at the midpoint simultaneously.
Although chromatic dispersion always results in photon deformation, the indistinguishability will not be affected if both photons are subjected to the same amount of dispersion.
This is the consequence of a phenomenon known as dispersion cancellation \cite{fan2021, im2021a}.
The situation is different in case the midpoint station is placed asymmetrically.
If the photons travel through fibers of different lengths they will undergo different amounts of dispersion and hence be deformed differently.
\\

A wave packet $\phi$ in a one-dimensional medium emitted at time $t=t_0$ and location $x=0$ takes the form
\begin{equation}
\phi (x, t) =  \int d\omega \phi(\omega) e^{i \omega (t-t_0) - i \beta(\omega) x}.
\end{equation}
Here, $\beta(\omega)$ is the wave number corresponding to a monochromatic wave with angular frequency $\omega$,
which is determined by the medium the wave travels in.
Now, let $\phi_l$ ($\phi_r$) be the wave packet of the photon emitted by the left (right) node.
The indistinguishability $V$ between these photons at the midpoint station (i.e., at $x = L_\text{left}$ for $\phi_l$ and at $x = L_\text{right}$ for $\phi_r$) is then defined by
\begin{equation}
V = |\mu|^2,
\end{equation}
where $\mu$ is given by
\begin{equation}
\begin{aligned}
\mu &= \int d t \phi_l(L_\text{left}, t) \phi_r^*(L_\text{right}, t) \\
&= \int d \omega \phi_l(\omega) \phi_r^*(\omega) e^{i\beta(\omega) \Delta L + i \omega \Delta t}.
\end{aligned}
\end{equation}
Here, we have $\Delta t = t_l - t_r$, where $t_l$ ($t_r$) is the time of emission of the photon at the left (right) node.
As discussed in Section \ref{sec:imbalanced_losses}, the indistinguishability $V$ affects both the success probability and the fidelity of the single- and double-click protocols.
\\

We assume that the wave packets have a central frequency that is close to some frequency $\omega_0$.
It is then useful to Taylor expand the wave number of the fiber as \cite{mitschke2010}
\begin{equation}
\beta(\omega) \approx \beta_0 + \beta_1 (\omega - \omega_0) + \frac 1 2 \beta_2 (\omega - \omega_0)^2 + \frac 1 6 \beta_3 (\omega - \omega_0)^3 .
\end{equation}
Here, $\beta_0 = 1 / v_p$ and $\beta_1 = 1 / v_g$, where $v_p$ and $v_g$ are the phase and group velocity in the fiber respectively.
$\beta_2$ is the Group-Velocity Dispersion (GVD) parameter and $\beta_3$ the Third-Order Dispersion (TOD) parameter.
As the $\beta_0$ contribution will only alter the global phase of $\mu$, it does not affect the indistinguishability and can effectively be dropped from the expression.
Furthermore, we assume $\Delta t = - \beta_1 \Delta L + \delta t$, where $\delta t$ is the alignment mismatch (for $\delta t = 0$, both emissions are timed such that the photons arrive at the midpoint station exactly at the same time).
Then, using $\Delta \omega \equiv \omega - \omega_0$, we can effectively write
\begin{equation}
\label{eq:visibility_general}
\begin{aligned}
\mu =& \int \phi_l(\omega_0 + \Delta \omega) \phi_r^*(\omega_0 + \Delta \omega)\\
&\times e^{i\Delta L(\frac 1 2 \beta_2 \Delta \omega^2 +\frac 1 6 \beta_3 \Delta \omega^3) + i \delta t \Delta \omega}d \Delta \omega.
\end{aligned}
\end{equation}

The value of $V$ and how much it is degraded by chromatic dispersion depends on the exact shapes of the photons, i.e., on $\phi_l$ and $\phi_r$.
In general, we expect the photons to be affected by chromatic dispersion less if their spread in frequency is small, as frequency components that are far apart also travel at velocities that are far apart.
Below, we derive expressions for $V$ in case of two specific wave-packet shapes, namely Gaussian and Lorentzian.
(Attenuated) laser pulses are often approximated as Gaussian, and approximate Gaussian photons can, e.g., be produced using cavity quantum electrodynamics \cite{utsugi2022} or spontaneous four-wave mixing \cite{li2019a}.
We here take Gaussian wave packets as a generic example of a pulse which is well localised in time and frequency, allowing us to obtain analytical results.
On the other hand, Lorentzian photons are created through the radiative decay of a two-level system.
In practice, photons will rarely be exactly Gaussian or Lorentzian as they interact with other components in the system such as filters and cavities.
Yet, we can think of the two types of photons as two extremes in how spread out their frequency distributions are, and therefore how sensitive they are to chromatic dispersion.
It was noted in Ref. \cite{rohde2005} that a Gaussian wave packet, for a fixed value of the time-distribution standard deviation, has a frequency-distribution standard deviation that is as small as possibly allowed by the Heisenberg uncertainty principle.
From this the authors concluded that Gaussian photons offer the best protection against alignment mismatch $\delta t$.
Here, it leads us to expect Gaussian photons are well protected against chromatic dispersion.
Lorentzian photons on the other hand have frequency distributions with very long tails, with $|\phi_{l/r}(\omega)|^2$ only going to zero as $\tfrac 1 {\omega^2}$.
It is expected that they are therefore much more susceptible to the effects of chromatic dispersion.

\subsubsection{Gaussian photons}

The wave packets of two Gaussian photons with frequency mismatch $\delta \omega$ can be written as
\begin{equation}
\begin{aligned}
\label{eq:Gaussian_wavepacket}
\phi_{l/r}(\omega) = \frac 1 {\sqrt[4]{2\pi \sigma^2}} e^{- \frac 1 {4 \sigma^2} (\omega - \omega_0 \pm \frac 1 2 \delta \omega)^2}. \\
\end{aligned}
\end{equation}
The probability distributions $|\phi_{l/r}(\omega)|^2$ are Gaussian with standard deviation $\sigma$.
When there is no TOD, the indistinguishability can be calculated exactly, giving
\begin{equation}
\label{eq:Gaussian_exact_no_TOD}
V|_{\beta_3=0} = \frac {\exp\left(- 2 \left(\frac {\delta \omega}{\sigma}\right)^2 - \frac{(\delta t \sigma)^2}{1 + \Delta L^2 \beta_2^2 \sigma^4} \right)} {\sqrt{1 + \Delta L^2 \beta_2^2 \sigma^4}}.
\end{equation}
We derive this result in Appendix \ref{app:V}.
A similar expression has been derived under the more restrictive assumption $\delta t = \delta \omega = \beta_3 = 0$ in Ref. \cite{kambs2019}, with which ours is consistent.
In case the photon indistinguishability is close to one, $1 - V|_{\beta_3 = 0} \ll 1$, it is well-approximated by the leading-order expansion
\begin{equation}
\label{eq:Gaussian_approx_no_TOD}
V|_{\beta_3=0} \approx 1 - 2 \left(\frac {\delta \omega}{\sigma}\right)^2 - (\delta t \sigma)^2 - \frac 1 2 \Delta L^2 \beta_2^2 \sigma^4.
\end{equation}
Finding an exact solution to Equation \eqref{eq:visibility_general} when the TOD is nonzero is difficult, but a leading-order result can be readily found to yield
\begin{equation}
\begin{aligned}
\label{eq:Gaussian_V}
V =& V|_{\beta_3=0} \left( 1 - \Delta L \beta_3 \delta t \sigma^4 \right) \\
&+ \mathcal O \left( \Delta L^2 \beta_3^2 \sigma^6, \Delta L^3 \beta_2^2 \beta_3 \delta t \sigma^8, \Delta L \beta_3 \delta t^3 \sigma^6  \right).
\end{aligned}
\end{equation}
This result as well is derived in Appendix \ref{app:V}.
Note that, to first order, the TOD does not affect the indistinguishability in case $\delta t = 0$.
If the alignment mismatch is itself small, $|\delta t \sigma| \ll 1$, we can expect the TOD to have only a very small effect on the indistinguishability.

\subsubsection{Lorentzian photons}

For two Lorentzian wave packets with frequency mismatch $\delta \omega$ we can write
\begin{equation}
\begin{aligned}
\label{eq:Lorentzian_wavepacket}
\phi_{l/r}(\omega) &= \sqrt \frac{2\tau}{\pi} \frac{1}{1 - 2i\tau(\omega - \omega_0 \pm \frac 1 2 \delta \omega)}.\\
\end{aligned}
\end{equation}
While the corresponding frequency distributions are Lorentzian functions with $\tfrac 1 \tau$ as full width at half maximum, the time distributions of these photons are one-sided exponentials with standard deviation $\tau$.
We are not aware of an analytical method for determining the indistinguishability for Lorentzian photons in full generality.
One method to evaluate the indistinguishability is numerical integration as done in Refs. \cite{vural2018, weber2019}.
Instead we make the simplifying assumptions that the photons arrive at the same time ($\delta t = 0$), they have the same central frequency ($\delta \omega = 0$), and there is no TOD ($\beta_3=0$).
The indistinguishability then becomes exactly solvable, giving (see Appendix \ref{app:V} for a derivation)
\begin{equation}
\label{eq:Lorentzian_V}
V|_{\delta t = \delta \omega = \beta_3 = 0} = 1 - \frac{2 \sqrt 2}{\sqrt \pi} (C + S) + \frac 4 \pi (C^2 + S^2).
\end{equation}
Here, $C$ and $S$ are Fresnel integrals defined by $S = \int_0^x \sin(t^2) dt$ and $C = \int_0^x \cos(t^2) dt$ with $x = \sqrt{\tfrac 1 2 |\Delta L \beta_2| \tau^{-2}}$.
To linear order, $C = x$ and $S = 0$, and therefore when the effect of dispersion is small we can use the approximation
\begin{equation}
\label{eq:Lorentzian_V}
V|_{\delta t = \delta \omega = \beta_3 = 0} = 1 - \frac 2 {\sqrt \pi} \frac {\sqrt{|\Delta L \beta_2|}}{ \tau} + \mathcal O (|\Delta L \beta_2| \tau^{-2}).
\end{equation}
We stress that the assumption $\delta t = \delta \omega = \beta_3 = 0$ is not generally expected to hold in a real experiment; it is introduced solely to make the problem analytically more tractable.
However, by comparing to results obtained through numerical integration we find that the assumption $\beta_3 = 0$ does not greatly affect the result in conditions typical to single-mode fiber (see the discussion below and Figure \ref{fig:dispersion_numerical}).
Therefore, while the above equations may not be able to capture the effects of $\delta t$ and $\delta \omega$, they do accurately capture the effects of asymmetry in the placement of the midpoint station, as is the focus of this section.
Furthermore, we note that it may sometimes already be desirable to use frequency conversion to convert photons to frequencies that incur relatively less attenuation losses in fiber.
This opens up the possibility for correcting any frequency mismatch and bringing $\delta \omega$ close to zero \cite{stolk2022}.

\subsubsection{Requirements for indistinguishable photons}

\begin{figure}[h]
    \includegraphics[width=\linewidth]{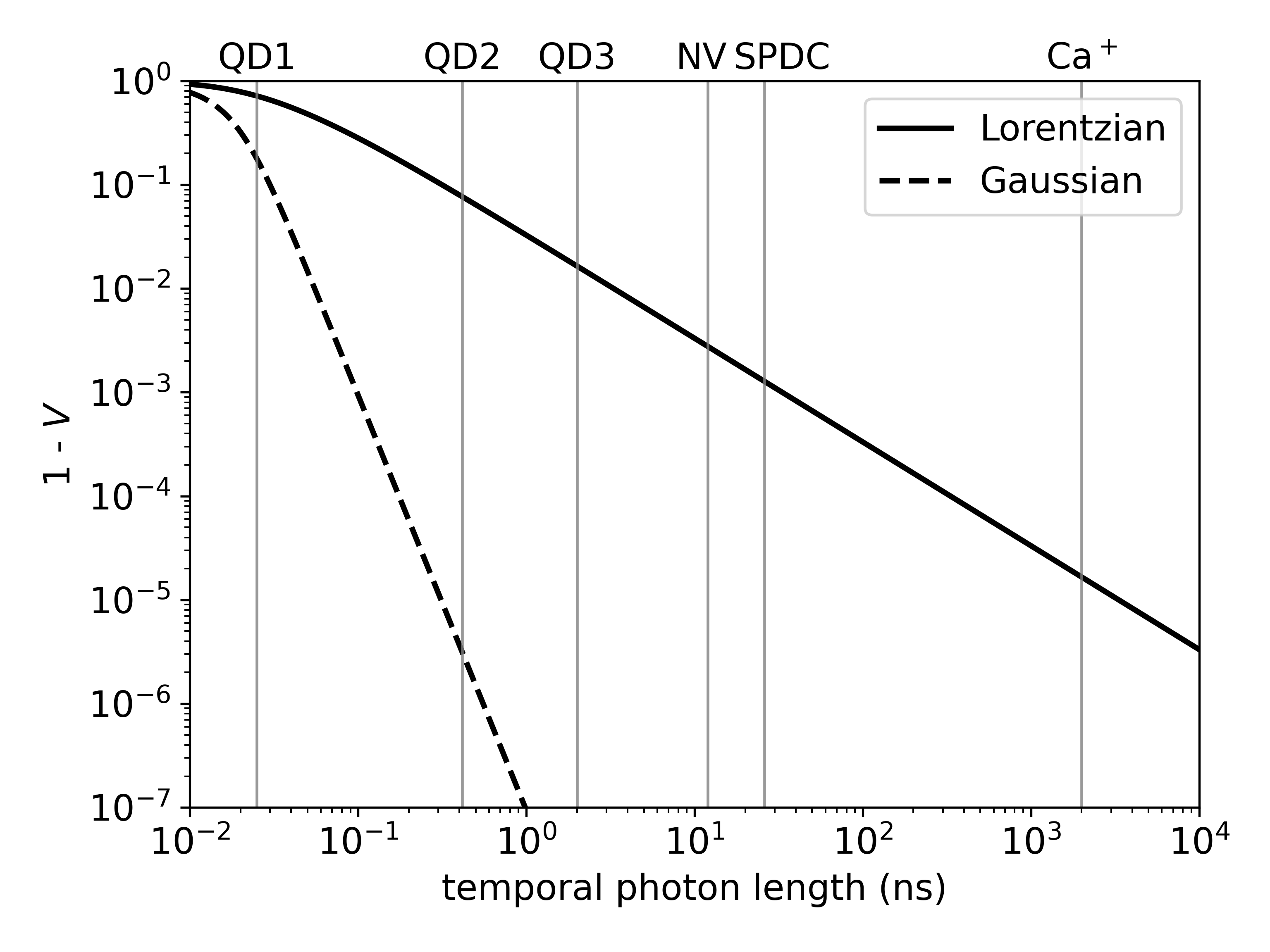}
    \caption[
        Comparison of visibilities of Gaussian and Lorentzian photons as function of the photon length.
        ]
    {
        Loss in the indistinguishability $V$ as a function of the temporal photon length, as measured by the standard deviation of the time distribution $|\phi(x=0, t)|^2$.
        We include results for both Gaussian and Lorentzian photons (Equations \eqref{eq:Gaussian_V} and \eqref{eq:Lorentzian_V}), for which the standard deviations are given by $\tfrac{1}{\sqrt 2 \sigma}$ and $\tau$ respectively.
        The results assume $\Delta L = 40$ km and a GVD of $\beta_2 \approx -21.7 \text{ps}^2/\text{km}$ (corresponding to a dispersion coefficient of 17 ps/(nm km), which is a typical value for single-mode optical fiber at 1550 nm \cite{zotero-4083}).
        The TOD parameter has been set to $\beta_3 = 0$.
        Other sources of noise are not included.
        That is, $\delta t = \delta \omega = 0$.
        The lengths of photons emitted by some specific sources have been indicated in the figure.
        QD1, QD2, QD3: quantum-dot sources from Refs. \cite{rota2022}, \cite{he2013} and \cite{schnauber2019} respectively.
        NV: nitrogen-vacancy centers in diamond \cite{goldman2015, kalb2018a, stolk2022}.
        (Some types of trapped ions, such as $\text{Ba}^+$ \cite{crocker2019} and $\text{Sr}^+$ \cite{stephenson2020} emit photons at a length close to the NV one.)
        SPDC: frequency-multiplexed spontaneous parametric down-conversion sources that interface with atomic quantum memories \cite{chakraborty2022, businger2022}.
        $\text{Ca}^+$: trapped calcium ions \cite{krutyanskiy2023} (lifetime estimated in \cite{avis2022a}).
    }
    \label{fig:dispersion}
\end{figure}

\begin{figure}[h]
    \includegraphics[width=\linewidth]{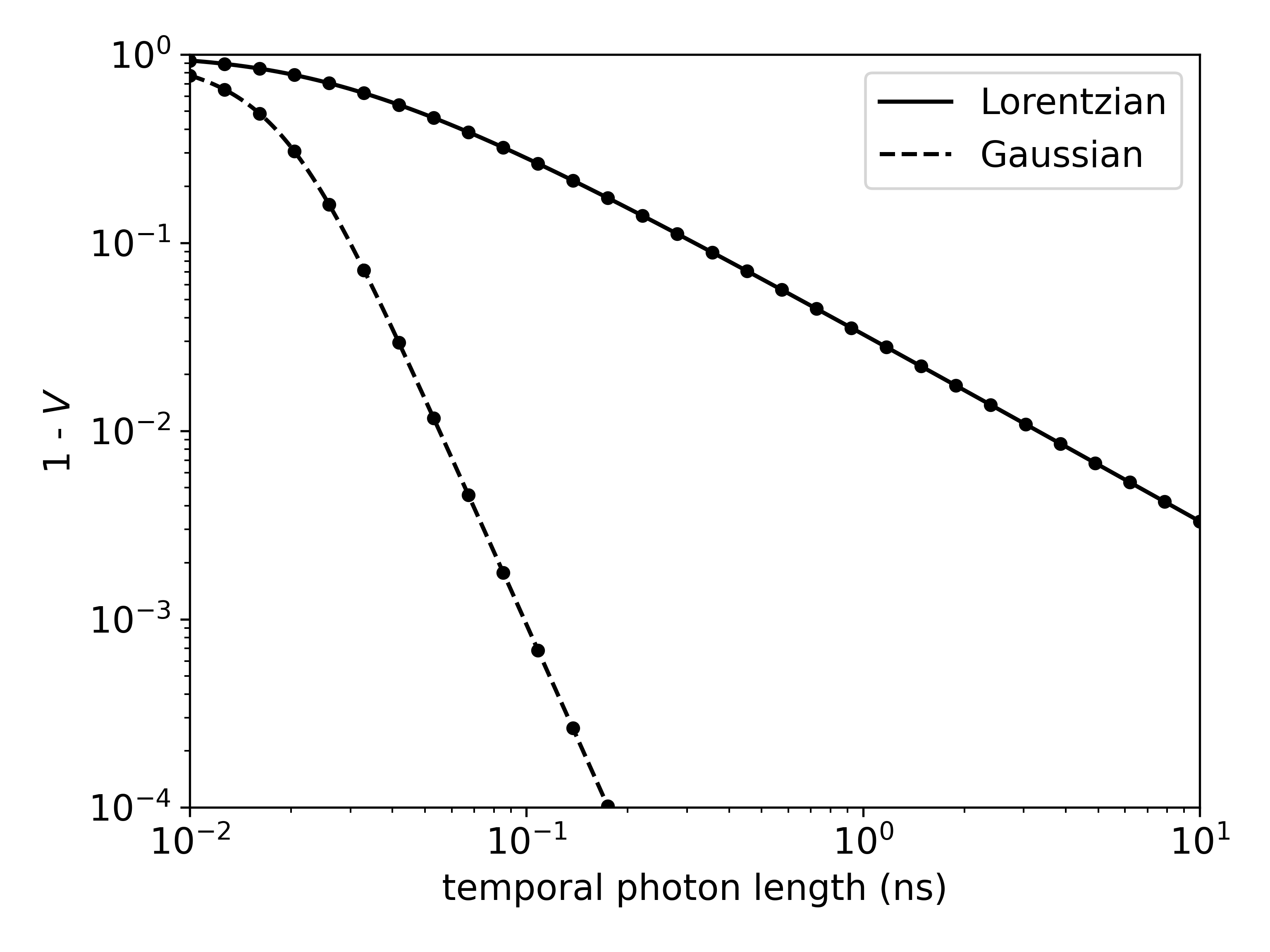}
    \caption[
        Comparison of analytical results to numerical results.
        ]
    {
        A comparison between our analytical results (the lines) for the photon indistinguishability $V$ assuming the TOD is zero (Equations \eqref{eq:Gaussian_exact_no_TOD} and \eqref{eq:Lorentzian_V}) and results obtained through numerical integration assuming a nonzero TOD (the markers).
        The temporal photon length is used as x axis, as measured by the standard deviation of $|\phi(x=0, t)|^2$.
        The results assume $\Delta L = 40$ km, a GVD of $\beta_2 \approx -21.7 \text{ps}^2/\text{km}$ and a TOD parameter of $\beta_3 \approx 0.127 \text{ps}^3 / \text{km}$ (corresponding to a dispersion coefficient of 17 ps/(nm km) and a dispersion slope of 0.056 ps/(nm$^2$ km), which are typical values for single-mode optical fiber at 1550 nm \cite{zotero-4083}).
        Other sources of noise are not included.
        That is, $\delta t = \delta \omega = 0$).
        Error bars for the numerical results are smaller than the marker size.
    }
    \label{fig:dispersion_numerical}
\end{figure}

The results above describe how the indistinguishability $V$ is diminished through the effect of chromatic dispersion in the case of asymmetric midpoint placement.
From these results, it becomes clear that how badly $V$ is reduced depends on the characteristics of the photon.
In particular, for Gaussian photons it depends on the parameter $\sigma$, while for Lorentzian photons it depends on the parameter $\tau$.
As expected, for both photons the effect of dispersion is increased as the width of the frequency distribution is increased, or equivalently, as the length of the time distribution is decreased.
In Figure \ref{fig:dispersion} we investigate how much indistinguishability is lost as a function of the length of the photon wave packet, assuming the photons are otherwise perfectly indistinguishable.
We here make the simplifying assumption that there is no TOD, thereby enabling the use of the exact analytical results obtained above.
This simplification is motivated by the fact that comparing our analytical results in case the TOD is zero with results obtained through numerical integration for a typical value of the TOD in single-mode optical fiber suggests that the TOD has only a negligible effect in this case, as shown in Figure \ref{fig:dispersion_numerical}.
Unsurprisingly we see in Figure \ref{fig:dispersion} that Lorentzian photons with their long tails in frequency are affected (much) worse by chromatic dispersion than Gaussian photons with the same length.
However, even for Lorentzian photons we see that (for standard single-mode fiber and $\Delta L = 40$ km) the decrease in $V$ is only of the order $10^{-2}$ when the length of the wave packet is of the order of nanoseconds.
\\

In case the photon length and $\Delta L$ are such that the decrease in photon indistinguishability can be significant,
it is clear that it is better if the photons are closer in shape to a Gaussian than a Lorentzian.
This strengthens the case for Gaussian photons made in Ref. \cite{rohde2005}, where it was found that Gaussian photons protect favourably against alignment mismatch.
However, some sources naturally emit photons that are more Lorentzian than Gaussian.
Potentially, photon-shaping techniques could be used to convert such photons to a more Gaussian waveform \cite{fioretto2020, keller2004, knall2022, morin2019, nisbet-jones2011}.
A simpler solution could be to send Lorentzian photons through a filter to remove the long tails of their frequency distribution.
While this would introduce extra losses, the spread in frequency could be greatly reduced, resulting in a much more Gaussian photon.
\\

Lastly we point out that there are various methods for reducing the drop in indistinguishability in case of asymmetric midpoint placement irrespective of photon shape.
The telecom C-band (1530 nm - 1565 nm) is the band conventionally used to transmit signals as it minimizes attenuation losses (a typical value of 0.275 dB/km in standard single-mode fiber \cite{zotero-4083}).
In contrast the telecom O-band (1260 nm - 1360 nm) incurs much heavier attenuation losses (typically 0.5 dB/km \cite{zotero-4083}), but as it is centered around the zero-dispersion wavelength (1310 nm) of standard single-mode optical fiber it minimizes dispersive effects.
By using the O-band instead of the C-band one can lessen the effects of chromatic dispersion at the cost of incurring extra losses.
This strategy is utilized in e.g., Ref.~\cite{shi2020}.
An investigation in Ref.~\cite{kambs2019} based on Gaussian photons suggests that using the O-band may only be worth it for photons shorter than approximately 100 picoseconds.
A second potential solution is the use of dispersion-shifted fiber.
Such fiber has its zero-dispersion wavelength in the telecom C-band and provides simultaneously small dispersion and small attenuation loss \cite{zotero-4143}.
However, such fiber is not widely deployed \cite{kambs2019, fasel2004} and hence not suitable when using existing fiber infrastructure to build a quantum network \cite{rabbie2022}.
Finally, one can use dispersion-compensating modules to reduce the effects of chromatic dispersion at the cost of incurring extra losses \cite{fasel2004}.

\section{Asymmetry in repeater chains}
\label{sec:repeater}

Now we turn our attention away from the placement of midpoint stations and instead consider the placement of repeater nodes in a quantum-repeater chain.
First, in Section \ref{sec:repeater_type}, we discuss the specific type of quantum-repeater chains we consider here.
Then, we pose two research questions about asymmetry in such repeater chains in Section \ref{sec:repeater_questions}.
These questions are made more precise in Sections \ref{sec:repeater_quantifying_performance}, \ref{sec:repeater_quantifying_asymmetry}, and \ref{sec:repeater_model}.
This allows us to address the research questions using numerical simulations in Section \ref{sec:repeater_numerical}.
Finally, we reflect on the numerical results in Section \ref{sec:repeater_reflections}.

\subsection{SWAP-ASAP repeaters with parallel entanglement generation}
\label{sec:repeater_type}

While there exist many types of quantum repeaters \cite{inside_quantum_repeaters, azuma2022}, we here focus on one specific type, namely the processing-node quantum repeater.
Such quantum repeaters are capable of generating and storing entanglement with neighboring nodes and of executing quantum gates.
These gates allow processing nodes to perform deterministic entanglement swapping, which is an operation such that if one qubit is entangled with some qubit A and the other qubit is entangled with some qubit B, performing entanglement swapping on those two qubits will result in qubits A and B being entangled \cite{bennett1993}.
Various proposed repeater platforms are processing nodes, such as trapped ions \cite{krutyanskiy2023a, sangouard2009, santra2019a, dhara2022}, color centers in diamond \cite{pompili2021, ruf2021, rozpedek2019} and neutral atoms \cite{langenfeld2021a, reiserer2015a}. 
\\

We here assume that each repeater has exactly two qubits, each of which can be used in parallel to perform heralded entanglement generation (as discussed in Section \ref{sec:midpoint}) with a different neighboring node.
(Note that there exist also proposed repeater systems that can only generate entanglement with one neighbouring node at a time \cite{avis2022a, rozpedek2018, pompili2021}.)
A chain of such repeaters can then create end-to-end entanglement by combining heralded entanglement generation and entanglement swapping.
How these are combined exactly, and what additional operations are performed, is dictated by the protocol that the repeaters execute.
Examples of additional operations that could be included are discarding entangled states when they have been stored in memory for too long \cite{rozpedek2018, li2021, khatri2021a, inesta2023, haldar2023} and entanglement distillation \cite{bennett1996, bennett1996a, dur2007}, both of which can help mitigate the effects of noise.
Optimizing repeater protocols is by no means an easy matter, and what protocols perform well depends both on the specific hardware used and the performance metric employed \cite{jiang2007, goodenough2021, rozpedek2018a, krastanov2019, li2021, inesta2023, haldar2023}.
Here, we consider the SWAP-ASAP protocol, in which no additional operations are included.
In the SWAP-ASAP protocol, each pair of neighboring nodes performs entanglement generation whenever this is possible.
That is, whenever at each node the qubit that is reserved for entanglement generation along that specific link is free.
As soon as both qubits at a repeater node are entangled it performs entanglement swapping (thereby freeing both qubits up again).
We have chosen to study this protocol as it is relatively simple both to understand and to study numerically.
Moreover, it has been found that the SWAP-ASAP protocol outperforms schemes that include entanglement distillation for near-term hardware quality, as measured both by the fidelity of end-to-end entangled states and the generation duration \cite{coopmans2021}.
Additionally, for the case when entanglement swapping is deterministic and entanglement is never discarded, it was found that the SWAP-ASAP protocol results in an optimal generation duration \cite{inesta2023, haldar2023}.
Throughout the rest of this paper, it will be assumed that quantum repeaters can generate entanglement with two neighbours in parallel and that they execute the SWAP-ASAP protocol.

\subsection{Research questions}
\label{sec:repeater_questions}

Asymmetric node placement will result in some fiber links between repeaters being shorter while others are longer.
As attenuation losses grow exponentially with the fiber length,
the longer links generate entanglement at a slower rate, and the shorter links at a faster rate.
In other words, the entangling rates along the chain become uneven due to asymmetric repeater placement.
The slower links could then potentially become bottlenecks.
This is expected to increase not only the amount of time required to distribute end-to-end entanglement,
but also the amount of time entangled states need to be stored in the SWAP-ASAP quantum repeaters until entanglement swapping takes place.
The result of this would be an increased amount of noise due to memory decoherence.
\\

The above observation motivates posing the following research question: what is the effect of uneven entangling rates in a SWAP-ASAP repeater chain in which repeaters can generate entanglement with both neighboring nodes simultaneously, as caused by the asymmetric distribution of the repeater nodes, on the performance of that chain?
A particularly simple method that could perhaps be used to mitigate any negative effects of asymmetric node placement is what we here refer to as the ``extended-fiber method''.
In this method, spooled fiber is used at the repeater nodes to make the shorter links as long as the longer ones, thereby effectively making the repeater chain symmetric again.
However, rather than making the bottlenecks faster, this method just makes the faster links slower.
It seems perhaps unlikely that such a strategy can lead to any improvement.
Therefore, we pose a second research question: is the extended-fiber method effective at improving the performance of asymmetric SWAP-ASAP repeater chains in which repeaters can generate entanglement with both neighboring nodes simultaneously?
In order to address these questions, they need to be made more precise.
To that end, we first quantify how well a repeater chain performs in Section \ref{sec:repeater_quantifying_performance}.
Then, we quantify how asymmetrical a repeater chain is and how we can systematically vary the amount of asymmetry in Section \ref{sec:repeater_quantifying_asymmetry}.
Finally, we introduce a simplified model for repeater chains in Section \ref{sec:repeater_model}.
\\

\subsection{Quantifying repeater performance}
\label{sec:repeater_quantifying_performance}

We quantify the performance of a repeater chain in terms of how capable it is at supporting Quantum Key Distribution (QKD).
Specifically, we consider the rate at which a secret key can be obtained when executing an entanglement-based implementation of the BB84 protocol \cite{bennett1992a, bennett2014} in the asymptotic limit.
The end nodes realize this protocol by keeping entangled quantum states stored in memory until they learn that all required entanglement swaps have been performed and hence end-to-end entanglement has been created.
At that time, they each measure their qubit in either the Pauli X or Z basis.
The corresponding asymptotic secret-key rate is given by \cite{shor2000}
\begin{equation} \label{eq:secret_key_rate}
\text{SKR} = \frac 1 T \max(1 - 2h(Q), 0).
\end{equation}
Here, $T$ is the generation duration, i.e., the average time required to distribute an end-to-end entangled state, $Q$ is the Quantum-Bit Error Rate (QBER), and $h(x) = -x \log_2 (x) - (1-x) \log_2 (1-x)$ is the binary entropy function.
The QBER is defined as the probability that, if both end nodes measure their qubits in the same basis, the parity between the outcomes is different than would be expected for the maximally-entangled target state.
Therefore, the QBER can be considered a measure for the amount of noise.
Note that, in general, the QBER can take a different value for measurements in the X basis than in the Z basis.
However, as we will be using a depolarizing noise model (see Section \ref{sec:repeater_model} below), the two values will coincide.
Our choice for the secret-key rate as performance metric is motivated not only by the fact that it has a clear operational interpretation,
but also because it combines information about how quickly and how noisily entanglement is distributed into a single convenient number.
While the secret-key rate is the primary performance metric considered here, the generation duration and QBER from which the secret-key rate is calculated can help provide a more detailed understanding of a repeater chain's performance.
\\

\subsection{Quantifying chain asymmetry}
\label{sec:repeater_quantifying_asymmetry}

Now, we first discuss how asymmetry in a repeater chain can be quantified.
We then use that to introduce a specific method for placing repeaters in a chain in such a way that the amount of asymmetry can be varied.
Let $\mathcal R$ be the set of all repeater nodes in the chain of interest.
Then, for every $n \in \mathcal R$, the \textit{node asymmetry parameter} is defined by
\begin{equation}
\label{eq:node_asym_param}
\mathcal A_n = \frac{ | L_\text{left of $n$} - L_\text{right of $n$} | }{ L_\text{left of $n$} + L_\text{right of $n$} }
\end{equation}
and the \textit{node asymmetry sign} is defined by
\begin{equation}
\label{eq:node_asym_sign}
S_n = \text{Sgn}\left(L_\text{left of $n$} - L_\text{right of $n$}\right).
\end{equation}
Here, $L_\text{left of $n$}$ ($L_\text{right of $n$}$) is the fiber distance between node $n$ and its neighboring node to the left (right) and Sgn is the sign function.
We note that $\mathcal A_n$ is equivalent to $\Delta L / L_\text{tot}$ and $S_n$ to $\text{Sgn}(\Delta L)$, where $\Delta L$ and $L_\text{tot}$ are defined for that specific node as in Equation \eqref{eq:L}.
While $\Delta L$ proved convenient to describe the effects of asymmetry in the placement of midpoint stations, we find the node asymmetry parameter more convenient in the context of repeater chains.
This is because the value of $L_\text{tot}$ can vary between different nodes in the chain, making it hard to understand just how asymmetrically a node is placed between its neighboring nodes from only $\Delta L$.
The node asymmetry parameters and node asymmetry signs of all repeater nodes together provide a complete parameterization of the locations of the nodes in the chain.
Now, we define the \textit{chain asymmetry parameter} $\mathcal A_\text{chain}$ to be the average value of $\mathcal A_n$ over all repeaters,
\begin{equation}
\label{eq:chain_asym_param}
\mathcal A_\text{chain} = \frac 1 {|\mathcal R|} \sum_{n \in \mathcal R} \mathcal A_n.
\end{equation}
While the node asymmetry parameter $\mathcal A_n$ quantifies how asymmetrically one specific node is placed between its neighboring nodes, the chain asymmetry parameter $\mathcal A_\text{chain}$ aims to capture how asymmetric the chain is as whole.
\\

\begin{figure}[h]
    \includegraphics[width=\linewidth]{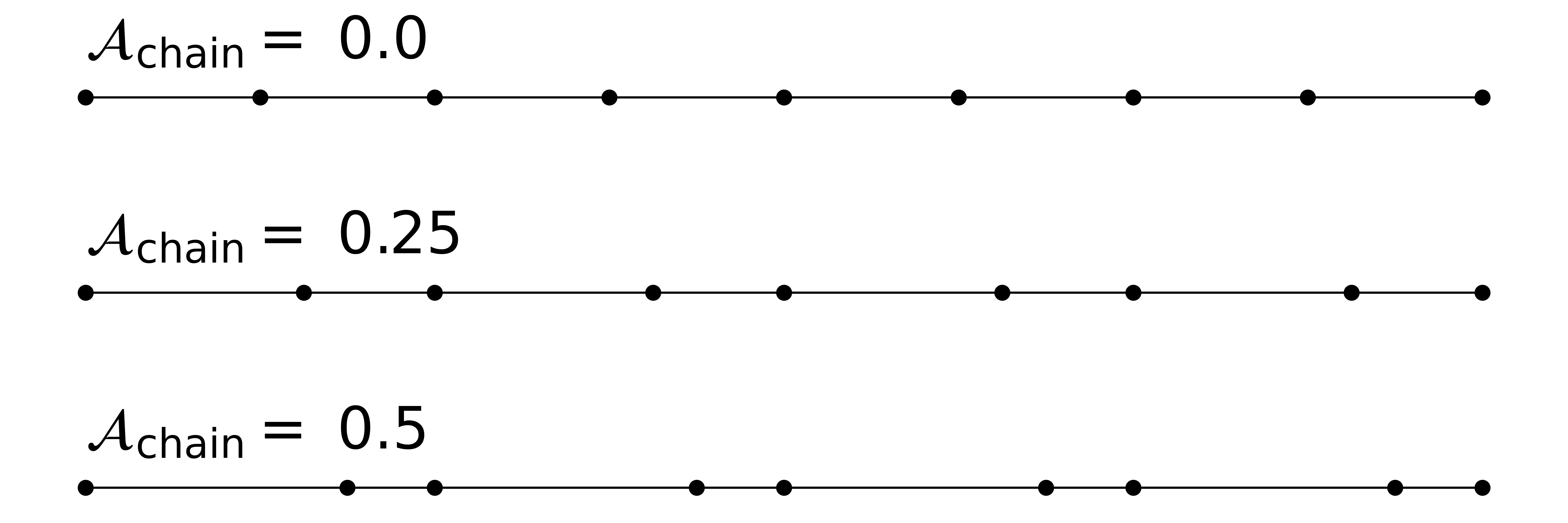}
    \caption[
        Locations of nodes in chain of varying asymmetry.
        ]
    {
        Locations of nodes in a chain with 7 repeaters for which $\mathcal A_n = \mathcal A_\text{chain}$ is the same for all nodes and $S_n$ alternates between nodes (see definitions in Equations \eqref{eq:node_asym_param}, \eqref{eq:node_asym_sign} and \eqref{eq:chain_asym_param}).
        The chain asymmetry parameter $\mathcal A_\text{chain}$ then quantifies the amount of asymmetry.
    }
    \label{fig:chain_node_locations}
\end{figure}

We aim to address the research questions posed in Section \ref{sec:repeater_questions} by investigating how the repeater-chain performance varies as a function of $\mathcal A_\text{chain}$.
However, for a repeater chain with a given total length and given number of nodes, there are many different possible repeater placements for which the chain asymmetry parameter takes the same value.
Therefore, in order to avoid ambiguity, we here introduce a specific class of repeater chains for which the parameter $\mathcal A_\text{chain}$ (together with the total length and number of nodes) uniquely defines the locations of all the repeaters.
These are repeater chains for which $\mathcal A_n$ is the same for every repeater in the chain
and $S_n$ alternates between nodes (such that no two neighboring repeaters have the same sign).
It then holds that $\mathcal A_\text{chain} = \mathcal A_n$ for all $n \in \mathcal R$.
See Figure \ref{fig:chain_node_locations} for an example of what such a repeater chain looks like for different values of $\mathcal A_\text{chain}$.
Our reason for choosing this class of chains is that the chains are relatively regular and easy to understand, while at the same time increasing $\mathcal A_\text{chain}$ clearly increases the disparity between long and short links, allowing us to study the effect of different entangling rates between different nodes as we set out to do.
\\

\subsection{Model for Repeater Chain}
\label{sec:repeater_model}

We consider a simplified model for the repeater nodes as well as for heralded entanglement generation between neighboring nodes.
In this model, the midpoint stations studied in Section \ref{sec:midpoint} are abstracted away, such that we can focus on the placement of the repeater nodes only.
We then take the cycle time for performing one attempt at generating an entangled state between two neighboring nodes to be given by
\begin{equation}
\label{eq:repeaters_t_cycle}
T_\text{cycle} = \frac L c,
\end{equation}
where $L$ is the distance between the neighboring nodes and $c$ is again the speed of light in fiber (here taken to be 200,000 km/s).
We note that this is equivalent to the cycle time when entanglement between neighboring nodes is generated using a symmetrically placed midpoint (see Equation \eqref{eq:cycle_time}).
We model the success probability of each attempt as
\begin{equation}
\label{eq:repeaters_p_succ}
P_\text{succ} = e^{- \frac {L} {L_\text{att}}},
\end{equation}
where $L_\text{att} \approx 22$ km is the attenuation length corresponding to attenuation losses of 0.2 dB/km.
This model has been chosen both for simplicity and for not being overly specific to one particular protocol for heralded entanglement generation.
It reflects the exponential scaling of the success probability common to both the double-click and single-click protocols, and also to protocols based on the direct transmission of photons between neighboring nodes \cite{bhaskar2020, langenfeld2021, lin2009} (assuming dark counts do not contribute significantly).
Therefore it is expected to adequately capture, at least on a qualitative level, how uneven entangling rates arise due to asymmetric node placement in repeater chains based on heralded entanglement generation.
\\

We model the states created by heralded entanglement generation to be noiseless.
More precisely, whenever an attempt is successful, a pure Bell state $\tfrac 1 {\sqrt 2} (\ket{00} + \ket{11})$ is created.
We consider the repeater nodes to be largely perfect devices at which entanglement swapping can be performed noiselessly and deterministically.
The only imperfection modeled at both repeater nodes and end nodes is that while qubits are stored in quantum memory, they undergo memory decoherence.
For simplicity, we model memory decoherence as depolarizing noise according to
\begin{equation}
\label{eq:depolar_memory}
\rho \to e^{-\frac t {T_\text{coh}}} \rho + (1 - e^{-\frac t {T_\text{coh}}}) \frac {\mathbb 1}{2}.
\end{equation}
Here $t$ is the storage time and $T_\text{coh}$ the coherence time, which we take to be one second here (as demonstrated with nitrogen-vacancy centers in Ref. \cite{bradley2022}).
Given these assumptions, noise in end-to-end entangled states produced by the repeater chain has only two sources.
The first of these is repeaters storing entangled states in quantum memory until entanglement swapping takes place.
The second is end nodes storing entangled states until all entanglement swaps have been completed and the measurements required by the BB84 protocol are performed.
These are exactly the sources of noise that may be affected by uneven entangling rates in a repeater chain.
\\

\subsection{Numerical results}
\label{sec:repeater_numerical}

\begin{figure}[h]
    \includegraphics[width=.8\linewidth]{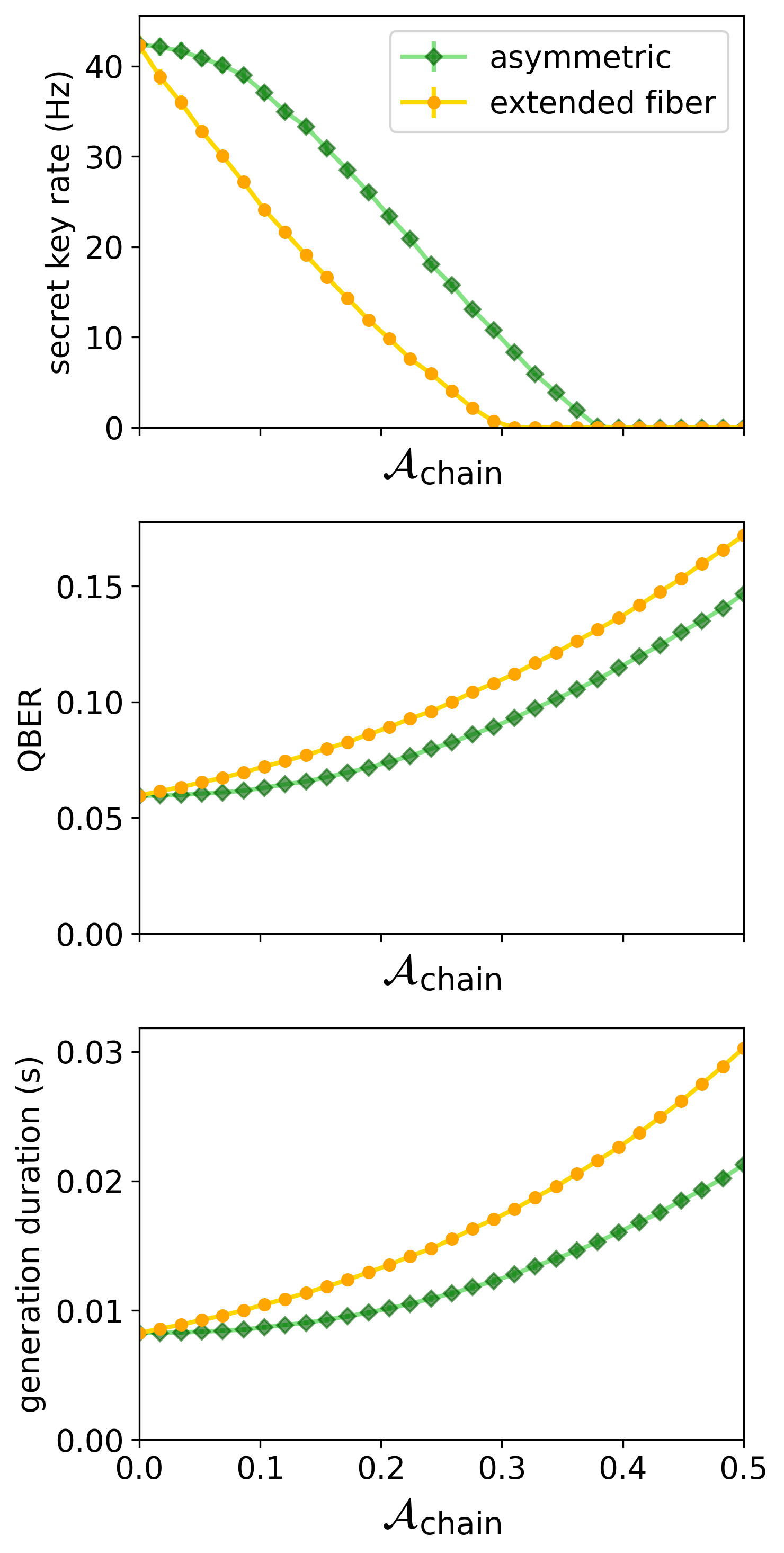}
    \caption[
        Effect of asymmetry in repeater chain on secret-key rate.
        ]
    {
        Effect of the chain asymmetry parameter $\mathcal A_\text{chain}$ in a repeater chain of the type illustrated in Figure \ref{fig:chain_node_locations} on the asymptotic secret-key rate of entanglement-based BB84.
        Additionally, the QBER and average entanglement-generation duration are shown, from which the secret-key rate is derived according to Equation \eqref{eq:secret_key_rate}.
        When using the ``extended-fiber method'', spooled fiber is deployed to make all links in the network equally long, resulting in an effectively symmetric network with an increased total fiber length.
        The total length of the repeater chain considered here is 1000 km and it contains 21 nodes (including 2 end nodes).
        Depolarizing memory decoherence (see Equation \eqref{eq:depolar_memory}) with a coherence time of $T_\text{coh}=1$s is the only source of noise included.
        Error bars represent the standard error in the estimates and are smaller than the marker size.
        Each data point is based on 20,000 simulated end-to-end entangled states.
    }
    \label{fig:repchain}
\end{figure}

Now, we are ready to address the research questions outlined in Section \ref{sec:repeater_questions}.
For concreteness, we consider a repeater chain with a length of 1000 km that contains 21 nodes (including two end nodes).
The nodes are thus, in the symmetric case, spaced 50 km apart.
A distance of 1000 km can be thought of as a typical pan-continental one, corresponding to, e.g.,  roughly the distance between Paris and Berlin.
In order to estimate the values of the generation duration and the QBER, we employ numerical simulations using the quantum-network simulator NetSquid \cite{coopmans2021}.
These simulations are based on the code introduced in Ref. \cite{avis2022a} and make use of a number of open-source libraries \cite{netsquid-magic, netsquid-netconf, netsquid-qrepchain, netsquid-simulationtools, netsquid-driver, netsquid-entanglementtracker, netsquid-physlayer}.
All simulation code and data can be found in our repository \cite{asymmetry_code}.
After the generation duration and QBER have been estimated, an estimate for the secret-key rate is computed using Equation \eqref{eq:secret_key_rate}.
The simulations are performed for different values of the chain asymmetry parameter $\mathcal A_\text{chain}$, and both for asymmetric chains and chains that have been symmetrized again using the extended-fiber method.
The results of these simulations are shown in Figure \ref{fig:repchain}.
It can be directly seen that using the extended-fiber method does not improve the performance of the repeater chain, but instead reduces it significantly.
Simulation results demonstrating that the same conclusion holds for other numbers of repeaters, other chain lengths and other coherence times can be found in our repository \cite{asymmetry_code}.
This suggests that the question whether the extended-fiber method can be used to mitigate the adverse effects of uneven entangling rates due to asymmetric repeater placement must be answered in the negative.
\\

It can be observed that the performance of the repeater chain exhibits some resilience against small amounts of asymmetry.
At $\mathcal A_\text{chain}=0.1$ the secret-key rate has only fallen by about 10\%.
At $\mathcal A_\text{chain}=0.20$,
corresponding approximately to Deutsche Telekom's planned trusted-node chain between Bonn and Berlin~\cite{dasilva2023},
it has only fallen by about 50\%.
For the specific repeater chain we consider, this corresponds to all the even nodes in the chain being displaced by 5 km and 10 km respectively as compared to their position in a symmetric chain, while the odd nodes remain in place (see also Figure \ref{fig:chain_node_locations}).
This resilience seems to be a consequence of the fact that both the generation duration and the QBER have a vanishing first derivative at $\mathcal A _\text{chain}=0$ in Figure \ref{fig:repchain}, in contrast to what is observed for the extended-fiber method.
We note furthermore that the first derivatives not only appear to vanish for the parameters considered in Figure \ref{fig:repchain}, but also for different numbers of nodes, chain lengths and coherence times, as demonstrated by data that can be found in our repository \cite{asymmetry_code}.
\\

\subsection{Reflection on numerical results}
\label{sec:repeater_reflections}

It may be surprising that the first derivatives in Figure \ref{fig:repchain} appear to vanish.
After all, when $\mathcal A_n$ is nonzero, the resultant longer links may be expected to form bottlenecks.
However, we need to take into account that while the longer links become slower at generating entanglement, the shorter links become faster.
It would appear that for small values of $\mathcal A_n$ the negative effect of the slower links is mostly compensated by the positive effect of the faster links.
We note that the same is not true when using the extended-fiber method, as in that case all links become slower due to asymmetry, and indeed the derivative does not appear to vanish at zero asymmetry in that case.
To foster an intuitive understanding, let us introduce the following hand-waving argument that reinforces the interpretation that first-order effects on the fast and slow links cancel each other.
Consider a single repeater node $n \in \mathcal R$.
This repeater is connected to its two neighbors by fibers of lengths $\tfrac 1 2 L_\text{tot} (1 \pm \mathcal A_n)$, where $L_\text{tot}$ is the sum of the two lengths.
Therefore, from combining Equations \eqref{eq:rate}, \eqref{eq:repeaters_t_cycle} and \eqref{eq:repeaters_p_succ}, we find that the average rates at which entanglement is generated with the two different neighbors are
\begin{equation}
\begin{aligned}
R_\pm &= \frac{ c\exp \left( - \frac{L_\text{tot}}{2 L_\text{att}} (1 \pm \mathcal A_n) \right)} {L_\text{tot} (1 \pm \mathcal A_n) }  \\
&= \frac {c e^{- \frac {L_\text{tot}}{2L_\text{att}}}} {L_\text{tot}} \left( 1 \mp (\frac{L_\text{tot}}{2L_\text{att}} + 1) \mathcal A_n \right) + \mathcal O (\mathcal A_n^2).
\end{aligned}
\end{equation}
Initially, entanglement generation is continuously attempted with both neighbors simultaneously.
This can be thought of as entanglement being created on one side with rate $R_+$ and with rate $R_-$ on the other side, resulting in a ``total rate'' at which entanglement is produced at this node of
\begin{equation}
R_\text{sum} = R_+ + R_- = 2 \frac {c e^{- \frac {L_\text{tot}}{2L_\text{att}}}} {L_\text{tot}}  + \mathcal O (\mathcal A_n^2).
\end{equation}
Abusively treating the time required to generate entanglement on either side as being exponentially distributed (while they are really geometrically distributed), we then have that the time required until the first entangled state is created takes time $1 / R_\text{sum}$.
This time is invariant with respect to the node asymmetry parameter at first order.
\\

Before entanglement swapping can take place, the second entangled state still needs to be generated.
Now, entangling attempts are only made on one side, and therefore the ``total rate'' is no longer $R_\text{sum}$ but only $R_+$ or $R_-$, depending on with which of the two neighbors entanglement has been established already.
The probability that the longer link is generated first (once more treating the times required to generate entanglement as being exponentially distributed) is given by $R_+ / R_\text{sum}$, in which case it on average still takes a time $1/R_-$ to generate the second entangled state.
Similarly, with probability $R_-/R_\text{sum}$ it still takes a time $1/R_+$.
Therefore, the average time until entanglement is swapped at repeater $n$ is
\begin{equation}
\begin{aligned}
T_\text{swap} &= \frac 1 {R_\text{sum}} \left( 1 + \frac{R_+}{R_-} + \frac{R_-}{R_+} \right) \\
&= \frac 3 2 \frac {c e^{- \frac {L_\text{tot}}{2L_\text{att}}}} {L_\text{tot}} + \mathcal O(\mathcal A_n^2),
\end{aligned}
\end{equation}
which is just the well-known ``three-over-two'' approximation for symmetric repeaters \cite{coopmans2022, jiang2007a}.
Furthermore, the average time during which the first entangled state is stored in quantum memory is then given by $T_\text{swap} - 1 / R_\text{sum}$, which also does not contain any linear terms in $\mathcal A_n$.
This is consistent with the fact that not only the generation duration of the repeater chain appears to be independent of the chain asymmetry parameter to linear order, but also the QBER.
\\

While the above argument can help understand why the performance of the repeater chain studied here has a vanishing first derivative with respect to the asymmetry parameter, we stress that it is not a complete or accurate treatment.
For one, we have approximated geometrically-distributed random variables as being exponentially distributed.
Moreover, we neglected the fact that in order to calculate the QBER we would need to calculate the expected value of the exponential function occurring in Equation \eqref{eq:depolar_memory}, which is not the same as the exponential function evaluated at the expected value.
But perhaps most importantly, the different repeaters cannot be considered in isolation.
After the repeater has performed entanglement swapping, it can only start entanglement generation again with neighbors that have themselves also performed entanglement swapping (otherwise their qubit is still occupied).
This complex interdependence is one of the main reasons why we have turned to numerical simulations here.

\section{Conclusion}

We have investigated how the asymmetric placement of nodes in a quantum network can affect network performance.
Specifically, we have studied the effect of asymmetric midpoint placement on heralded entanglement generation and of asymmetric repeater placement on SWAP-ASAP repeater chains in which repeaters can generate entanglement with both neighboring nodes in parallel.
In both cases we have observed a remarkable resilience against small amounts of asymmetry, even though performance can be expected to degrade significantly as asymmetry is increased further.
While for the midpoint placement the cycle time will be directly affected when asymmetry is introduced,
the success probability and fidelity have vanishing first derivatives.
Similarly, for repeater chains, both the generation duration and QBER appear to have vanishing first derivatives with respect to asymmetry in repeater placement.
Whether the first derivatives also vanish for repeater chains in which parallel entanglement generation is not possible remains an open question.
The same is true for repeater chains that do not execute a SWAP-ASAP protocol but instead, for example, execute a protocol that includes entanglement distillation.
Furthermore, we have assumed a specific form of asymmetric node placement (see Figure~\ref{fig:chain_node_locations}).
An interesting direction for further research is to compare these chains to real-world fiber infrastructure and investigate if our conclusions still hold when asymmetric chains are generated through different means, e.g., randomly.
\\

Our results suggest that extending fiber segments to symmetrize repeater placement is not an effective method for mitigating the effects of asymmetry.
An open question, then, is whether such methods exist.
One strategy, reminiscent of the extended-fiber method, could be to reduce the difference in the expected completion times of longer and shorter links by delaying entanglement generation on the short links.
However, as geometric distributions are memoryless, starting the shorter links at a delayed time when the longer links have not yet finished will result in the same waiting time as when the shorter link would have been started immediately.
But if the longer links had already finished at the delayed time, it would have been better if the shorter link had started at an earlier time.
Therefore we expect that fixed delays are not very effective, but dynamic delays where the shorter links are started depending on which links have succeeded already could work better and pose an interesting scheduling problem.
Another option is the use of cutoffs~\cite{rozpedek2018, li2021, khatri2021a, inesta2023, haldar2023}.
When using cutoffs, entangled qubits that have been stored in memory for too long are discarded to bound noise levels.
As cutoffs do not waste potentially useful resources by keeping links idle, we expect them to be more effective than delays.
Quantifying how well these various strategies mitigate the effects of asymmetry is an interesting avenue for further research.
\\

We have observed that asymmetry in midpoint placement can significantly affect the indistinguishability of photons used in heralded entanglement generation because of chromatic dispersion.
Chromatic dispersion can potentially cause a bad fidelity even for small amounts of asymmetry.
The size of the effect, however, depends on the temporal length of the photons, and we have found that as long as the photons are long enough (on the order of nanoseconds) the effect of chromatic dispersion can be negligible even for large asymmetries (percent level for an asymmetry of 40 km, see Figure \ref{fig:dispersion}).
We have furthermore found that Gaussian wave packets are much more resilient against chromatic dispersion than Lorentzian wave packets which have long tails in their frequency distribution.
By making the shape of a wave packet to be more Gaussian than Lorentzian (e.g., by filtering out long tails), the effects of chromatic dispersion can be mitigated.
\\

From all this, we conclude that while asymmetry degrades quantum-network performance and should therefore be avoided where possible, 
small amounts of asymmetry are not expected to have a large effect.
This may alleviate some of the pressure in selecting the perfect locations for nodes in a quantum network,
and suggests that at least those fiber networks that are only slightly asymmetric can provide fertile ground for a future quantum internet.

\section{Code availability}
The code that was used to perform the simulations and generate the plots in this paper has been made available at \url{https://gitlab.com/GuusAvis/reproduction-code-for-asymmetric-node-placement-in-fiber-based-quantum-networks} \cite{asymmetry_code}.

\section*{Acknowledgements}
We thank Francisco Ferreira da Silva, Janice van Dam, Arian Stolk and Kian van der Enden for useful discussions.
We thank Arian Stolk for critical reading of the manuscript.
This work was supported by the QIA-Phase 1 project which has received funding from the European Union’s Horizon Europe research and innovation programme under grant agreement No. 101102140, NWO Zwaartekracht QSC 024.003.037 and Danmarks Grundforskningsfond (DNRF 139, Hy-Q Center for Hybrid Quantum Networks).

\bibliography{asymmetry}

%%%%%%%%%%%%%%%%%%%%%%%%%%%%%%%%%%%%%%%%%%%%
%%%%%%%%%%%%% APPENDIX
%%%%%%%%%%%%%%%%%%%%%%%%%%%%%%%%%%%%%%%%%%%%

\appendix

\onecolumngrid

\section{Single-click and double-click expressions}
\label{app:single_double_click}

In this appendix we derive the success probability and fidelity of the single- and double-click protocols in terms of the parameter $\Delta L$ (to first order).
These derivations are based on the expressions given in the appendix of Ref. \cite{avis2022a}.
For both protocols, our derivation hinges on having a set of probabilities $\{p_i\}$ and a set of states $\{\rho_i\}$ such that
\begin{equation}
P_\text{succ} = \sum_i p_i
\end{equation}
is the success probability and
\begin{equation}
\rho = \frac 1 {P_\text{succ}} \sum_i p_i \rho_i 
\end{equation}
is the mixed state upon success.
The fidelity can then be written as
\begin{equation}
F = \frac 1 {P_\text{succ}} \sum_i p_i F_i
\end{equation}
where $F_i$ is the fidelity corresponding to $\rho_i$.
$p_i$ and $F_i$ depend on $P_\text{left}$ and $P_\text{right}$, which we then rewrite in terms of $P_\text{tot} = P_\text{left} P_\text{right}$ and $P_\text{sum} = P_\text{left} + P_\text{right}$ (see Equation \eqref{eq:p_tot_p_sum}).

\subsection{Double click}

For the double-click protocol we have (taking both the results and nomenclature from Ref. \cite{avis2022a})
\begin{equation}
\begin{aligned}
p_T =& \frac 1 2 P_\text{tot} V (1 - p_\text{dc}^{2r}), \\
p_{F_1} =& \frac 1 2 P_\text{tot} (1 - V) (1 - p_\text{dc}^{2r}), \\
p_{F_2} =&  \frac {2 - r}{2} P_\text{tot} (1 + V) p_\text{dc} (1 - p_\text{dc})^{r + 1}, \\
p_{F_3} =& 2 (P_\text{sum} - 2 P_\text{tot}) p_\text{dc} (1 - p_\text{dc})^{r+1}, \\
p_{F_4} =& 4 (1 - P_\text{sum} + P_\text{tot}) p_\text{dc}^2 (1 - p_\text{dc})^2,
\end{aligned}
\end{equation}
such that we have the sets $\{p_i\}$, $\{\rho_i\}$ and $\{F_i\}$ with 
$p_1 = q_\text{em} p_T$, 
$\rho_1 = \ketbra{\Psi^{\pm}}$
(where $\ket{\Psi^{\pm}} = \tfrac 1 {\sqrt 2} (\ket{01} \pm \ket{10})$ is the target Bell state, with the sign depending on which detector clicked),
$ F_1 = 1$, 
$ p_2 = q_\text{em} p_{F_1}$, 
$ \rho_2 = \frac 1 2 (\ketbra{01} + \ketbra{10})$, 
$ F_2 = \frac 1 2$, 
$ p_3 = q_\text{em} p_{F_2}$,
$ \rho_3 = \frac 1 2 (\ketbra{00} + \ketbra{11})$,
$ F_3 = 0$,
$ p_4 = (1 - q_\text{em}) (p_T + p_{F_1} + p_{F_2}) + p_{F_3} + p_{F_4}$,
$ \rho_4 = \frac{\mathbb 1}{4}$, 
and $ F_4 = \frac 1 4$.
From this, it follows that we can write
\begin{equation}
\begin{aligned}
P_\text{succ} =& a + b P_\text{sum}, \\
F =& \frac{c + \frac 1 4 b P_\text{sum}}{a + b P_\text{sum}},
\end{aligned}
\end{equation}
with
\begin{equation}
\begin{aligned}
a =& \frac 1 2 P_\text{tot} (1 - p_\text{dc})^{2r} + p_\text{dc} P_\text{tot} (1 - p_\text{dc})^{r+1} \left[\frac{2 - r} {2} (1 + V) - 4 \right]  + 4 p_\text{dc}^2 (1 + P_\text{tot})(1-p_\text{dc})^2, \\
b =& 2 p_\text{dc} (1 - p_\text{dc})^{r+1} - 4 p_\text{dc}^2 (1 - p_\text{dc})^2, \\
c =& \frac 1 4 q_\text{em} P_\text{tot}(1 + V) (1 - p_\text{dc})^{2r} + (1 - q_\text{em}) \left( \frac 1 8 P_\text{tot} (1 - p_\text{dc})^{2r} + \frac {2 - r} 8 p_\text{dc} (1 - p_\text{dc})^2  P_\text{tot} (1 + V) \right) \\
&- p_\text{dc} P_\text{tot} (1 - p_\text{dc})^{r+1}+ p_\text{dc}^2 (1 + P_\text{tot})(1-p_\text{dc})^2.
\end{aligned}
\end{equation}
Taking a first-order expansion in $p_\text{dc}$ of the success probability and the fidelity gives the double-click results presented in Section \ref{sec:imbalanced_losses}.

\subsection{Single click}

For the single-click protocol we can rewrite the expressions in Ref. \cite{avis2022a} by substituting the bright-state parameters by $q$ (see Equation \eqref{eq:single_click_q}) such that we have the sets $\{p_i\}$, $\{\rho_i\}$ and $\{F_i\}$ with
\begin{equation}
\begin{aligned}
p_1 =& \frac{q^2}{P_\text{tot}} (1 - p_\text{dc}) \left\{2 p_\text{dc} + P_\text{tot}\left(- 2 (1 - p_\text{dc})^{r-1} + 2 p_\text{dc} + \frac 1 2 (2 - r) (1 + V)) \right) \right\} \\
&+ q^2 \frac{P_\text{sum}}{P_\text{tot}} \bigg((1 - p_\text{dc})^{r} - 2 p_\text{dc} (1 - p_\text{dc})\bigg) \\
p_2 =& t_1 + t_2,\\
t_1 =& q\left(1 - q \frac  {P_\text{left}} {P_\text{tot}}\right) \left\{(1 - p_\text{dc})^r + 2 \left(\frac {P_\text{right}} {P_\text{tot}} - q \right)(1 - p_\text{dc})p_\text{dc} \right\}, \\
t_2 =& q\left(1 - q \frac  {P_\text{right}} {P_\text{tot}}\right) \left\{(1 - p_\text{dc})^r + 2 \left(\frac {P_\text{left}} {P_\text{tot}} - q \right)(1 - p_\text{dc})p_\text{dc} \right\}, \\
p_3 =& 2 p_\text{dc} (1 - p_\text{dc}) \left(1 + \frac {q^2}{P_\text{tot}} - q \frac{P_\text{sum}}{P_\text{tot}} \right) \\
\end{aligned}
\end{equation}
$\rho_1 = \ketbra{00}$, $\rho_3 = \ketbra {11}$,  $F_1 = F_3 = 0$,
\begin{equation}
\begin{aligned}
\rho_2 &= \frac 1 {p_2} \left( t_1 \ketbra{01} + t_2 \ketbra{10} \pm \sqrt{V t_1 t_2} (\ketbra{01} + \ketbra{10}) \right), \\
F_2 &= \frac 1 2 + \sqrt V \frac{ \sqrt{t_1 t_2} } {p_2}.
\end{aligned}
\end{equation}
We note that while $P_\text{left}$ and $P_\text{right}$ cannot be eliminated in favor of $P_\text{sum}$ and $P_\text{tot}$ in the expressions for $t_1$ and $t_2$, they can be eliminated in $p_2 = t_1 + t_2$, giving
\begin{equation}
\begin{aligned}
p_2 =& 2q(1 - p_\text{dc})^r - 4 q p_\text{dc} (1 - p_\text{dc}) (1 + \frac{q}{P_\text{tot}}) \\
& + \frac {P_\text{sum}} {P_\text{tot}} q \Bigg( 2p_\text{dc}(1 -p_\text{dc}) - q\bigg\{ (1 - p_\text{dc})^r - 2 (1 - p_\text{dc}) p_\text{dc}\bigg\} \Bigg)
\end{aligned}
\end{equation}
The success probability is then given by
\begin{equation}
P_\text{succ} = p_1 + p_2 + p_3.
\end{equation}
We refrain from writing out the exact success probability explicitly here, but note that it can be readily verified that the terms proportional to $P_\text{sum}$ cancel out.
Therefore, we conclude that the success probability is independent of $P_\text{sum}$, and hence the asymmetry.
To leading order we have $p_1 = 0$, $p_2 = 2q$ and $p_3 = 2 p_\text{dc}$, and by adding these up the leading-order result for the success probability in Equations \eqref{eq:single_click} is found.
\\

As both $F_1$ and $F_3$ are zero, the fidelity is given by
\begin{equation}
F = \frac{p_2}{P_\text{succ}} F_2 = \frac 1 2 \frac{p_2}{P_\text{succ}} + \sqrt V \frac{\sqrt{t_1 t_2}}{P_\text{succ}}.
\end{equation}
The product $t_1 t_2$ cannot be written in terms of solely $P_\text{tot}$ and $P_\text{sum}$ instead of $P_\text{left}$ and $P_\text{right}$.
However, the troublesome terms in this product are higher order.
Therefore, we can eliminate $P_\text{left}$ and $P_\text{right}$ from $F$ as long as we stick to leading order.
Here, we consider both $q$ and $p_\text{dc}$ to be of the same order, i.e., $p_\text{dc} = \mathcal O(q)$.
We note that under realistic settings $p_\text{dc} < q$ as otherwise more successes would be caused by dark counts than by actual photons, which is a regime in which no useful entanglement can be created.
Evaluating $F$ at leading order requires evaluating $p_2$, $P_\text{succ}$ and $\sqrt{t_1 t_2}$ up to second order, giving
\begin{equation}
\begin{aligned}
p_2 =& 2q \left( 1 - (2 + r) p_\text{dc} + \frac{P_\text{sum}}{P_\text{tot}} (p_\text{dc} - \frac 1 2 q) \right) + \mathcal O (q^3), \\
P_\text{succ} =& 2 (q + p_\text{dc}) \left( 1 - q - p_\text{dc} + \frac q {q + p_\text{dc}} \left[ \frac 1 4 (2 - r) (1 + V) q - 2 p_\text{dc} \right] \right) + \mathcal O(q^3), \\
\sqrt{t_1 t_2} =& q \left( 1 - (2 + r) p_\text{dc} - \frac 1 2 (q - 2 p_\text{dc}) \frac{P_\text{sum}}{P_\text{tot}} \right) + \mathcal O (q^3).
\end{aligned}
\end{equation}
The leading-order expression for the fidelity given in Equation \eqref{eq:single_click} can now be obtained by substituting these quantities into the equation for the fidelity above and disregarding higher-order terms.

\section{Photon Indistinguishability}
\label{app:V}

In this appendix we derive the formulas for indistinguishability of Gaussian and Lorentzian photons presented in the main text.

\subsection{Gaussian}

For the Gaussian wave packets given in Equation \eqref{eq:Gaussian_wavepacket}, $\mu$ as defined in Equation \eqref{eq:visibility_general} becomes
\begin{equation}
\mu = \frac 1 {\sqrt{2 \pi}}e^{-\left(\frac{\delta \omega}{\sigma}\right)^2} \int d x  \exp\left(i \delta t \sigma x - \frac 1 2 \left[1 - i \Delta L \beta_2 \sigma^2 \right] x^2 + \frac 1 6 i \Delta L \beta_3 \sigma^3 x^3\right).
\end{equation}
By Taylor expanding in the TOD parameter $\beta_3$ we can rewrite this as
\begin{equation}
\mu = \frac 1 {\sqrt{2 \pi}}e^{-\left(\frac{\delta \omega}{\sigma}\right)^2}  \sum_{n=0}^\infty \frac {i^n(\Delta L \beta_3 \sigma^3)^n} {n!6^n} \int d x x^{3n} \exp\left(i \delta t \sigma x - \frac 1 2 \left[1 - i \Delta L \beta_2 \sigma^2 \right] x^2 \right).
\end{equation}
This allows us to evaluate the indistinguishability at different orders of $\Delta L \beta_3 \sigma^3$ by evaluating the moments of a Gaussian distribution.
The $n = 0$ term, which corresponds to $\beta_3 = 0$, is a simple Gaussian integral for which it holds that
\begin{equation}
\label{eq:Gaussian_integral}
 \int_{-\infty}^\infty e^{-ax^2 + bx + c} dx = e^{\frac{b^2}{4a} + c} \sqrt{\frac{\pi}{a}}.
\end{equation}
Therefore,
\begin{equation}
\mu|_{\beta_3 = 0} = e^{-\left(\frac{\delta \omega}{\sigma}\right)^2} \frac 1 {\sqrt{ 1  - i \Delta L \beta_2 \sigma^2) }} e^{- \frac{(\sigma\delta t )^2}{2( 1  - i \Delta L \beta_2 \sigma^2)}}.
\end{equation}
The result in Equation \eqref{eq:Gaussian_exact_no_TOD} is then obtained through
\begin{equation}
V|_{\beta_3 = 0} = \Big| \mu|_{\beta_3=0} \Big|^2.
\end{equation}
To evaluate higher-order terms one can use
\begin{equation}
\int_{-\infty}^\infty x^n e^{-ax^2 + bx +c} dx = e^{\frac{b^2}{2a} + c} \sqrt{\frac \pi a}\left[\left(\frac{b}{2a}\right)^n + \sum_{j=1}^{\lfloor \frac n 2 \rfloor} \binom{2j}{n} \left( \frac  b {2a} \right)^{n-2j} \frac{(2j - 1)!!}{(2a)^j} \right].
\end{equation}
In fact, this allows for determining the indistinguishability to arbitrary order in the TOD.
Here, we only calculate the first order.
By recognizing that the expression before the square brackets in the equation above is equation to the result of the regular Gaussian integral, we can then write
\begin{equation}
\mu = \mu|_{\beta_3 = 0} \left(1 + i A \Delta L \beta_3 \sigma^3 \right) + \mathcal O \left( (\Delta L \beta_3 \sigma^3)^2\right)
\end{equation}
with
\begin{equation}
A = \frac{(i \delta t \sigma)^3 + 3 i \delta t \sigma (1 - i \Delta L \beta_2 \sigma^2) }{6 (1 - i\Delta L \beta_2 \sigma^2)^3} .
\end{equation}
We then find
\begin{equation}
V = |\mu|^2 = V|_{\beta_3=0} \left(1 + 2 \Delta L \beta_3 \sigma^3 \rm{Re}(iA) \right) + \mathcal O \left( (\Delta L \beta_3 \sigma^3)^2 \right).
\end{equation}
Evaluating this expression yields 
\begin{equation}
\begin{aligned}
V =& V|_{\beta_3=0} \Bigg[ 1 - \frac {\Delta L \beta_3 \delta t \sigma^4} {\left(1 + \Delta L^2 \beta_2^2 \sigma^4\right)^2}
\times \Bigg( 1 - 3 \Delta L^2 \beta_2^2 \sigma^4 \\
&- \frac{\delta t^2 \sigma^2 (1 - \Delta L^2 \beta_2^2 \sigma^4)^2}{3(1 + \Delta L^2 \beta_2^2 \sigma^4)} 
\Bigg) \Bigg] + \mathcal O (\Delta L^2 \beta_3^2 \sigma^6).
\end{aligned}
\end{equation}
Collecting the higher-order terms gives Equation \eqref{eq:Gaussian_V}.

\subsection{Lorentzian}

Using the Lorentzian wave forms defined in Equation \eqref{eq:Lorentzian_wavepacket} for $\delta \omega = \delta t = \beta_3 = 0$ we find
\begin{equation}
\mu = \frac 1 \pi \int_{-\infty}^\infty dx \frac{ e^{\pm i c x^2}}{1 + x^2}
\end{equation}
where $\pm c = \frac 1 2 \Delta L \beta_2 \tau^{-2}$ and $c>0$ is a real number (the sign $\pm$ is the sign of $\Delta L \beta_2$).
We rewrite this as
\begin{equation}
\mu = e^{\pm i c} I(\pm i c)
\end{equation}
with
\begin{equation}
I(y) = \int_{-\infty}^\infty dx \frac{e^{-y (1 + x^2)}}{1 + x^2}.
\end{equation}
We can evaluate this integral by first differentiating it and then integrating it again.
By the fundamental theorem of calculus, it holds that
\begin{equation}
I(y) = I(0) + \int_0^y \frac{d I(z)}{dz} dz.
\end{equation}
Both terms are readily evaluated;
\begin{equation}
I(0) = \int_{-\infty}^\infty \frac {dx}{1 + x^2} = \atan(x)|^\infty_{-\infty} = \pi,
\end{equation}
\begin{equation}
\begin{aligned}
\int_0^y \frac{d I(z)}{dz} dz =& \int_0^y dz \int_{-\infty}^\infty -e^{-z (1 + x^2)} dx \\
=& \int_0^y dz e^z \sqrt{ \frac{\pi}{z}} \\
=& -2 \sqrt{\pi} \int_0^{\sqrt{y}} e^{-u^2} du \\
=& - \pi \erf{\sqrt y},
\end{aligned}
\end{equation}
where $\erf$ is the error function.
Here we have assumed $\rm{Re}(z) \geq 0$ so that we could use Equation \eqref{eq:Gaussian_integral} (which is equivalent to assuming $\rm{Re}(y) \geq 0$) and we made a change of variables $u = \sqrt{y}$, where $\sqrt{y}$ is taken to mean the principal root of $y$.
Therefore, we have (for $\rm{Re}(y) \geq 0$)
\begin{equation}
I(y) = \pi (1 - \erf{\sqrt y}).
\end{equation}
In order to evaluate $\mu$ we need to evaluate the error function for a purely imaginary value.
In that case we can rewrite
\begin{equation}
\begin{aligned}
\int_0^{\sqrt{\pm i c} }e^{-u^2} du = \sqrt{\pm i} \int_0^{\sqrt c} e^{\mp i v^2} dv
= \frac{1}{\sqrt 2} \left( (1 \pm i) C(\sqrt{c}) + (1 \mp i) S(\sqrt c) \right).
\end{aligned}
\end{equation}
Here we made the change of variable $u = \sqrt{\pm i} v$ and we have introduced the Fresnel integrals $C(x) = \int_0^x \cos(t^2) dt$ and $S(x) = \int_0^x \sin(t^2) dt$.
Therefore, we have
\begin{equation}
I(\pm i c) = \pi - \sqrt{2 \pi} \left( (1 \pm i) C(\sqrt c) + (1 \mp i) S(\sqrt c) \right).
\end{equation}
We can then find the indistinguishability as given in Equation \eqref{eq:Lorentzian_V} as
\begin{equation}
V = |\mu|^2 = |I(\pm i c)|^2.
\end{equation}

\end{document}